\documentclass{tran-l}

\usepackage{epsfig}

\newtheorem{theorem}{Theorem}[section]
\newtheorem{lemma}[theorem]{Lemma}

\theoremstyle{definition}

\theoremstyle{remark}

\numberwithin{equation}{section}

\begin{document}

\title[Solving the KdV Equation by Its Bilinear Form]
{Solving the Korteweg-de Vries Equation by Its Bilinear Form:
Wronskian Solutions}

\author{Wen-Xiu Ma}
\address{Department of Mathematics,
University of South Florida, Tampa, FL 33620-5700}
\email{mawx@math.usf.edu}

\author{Yuncheng You}
\address{Department of Mathematics,
University of South Florida, Tampa, FL 33620-5700}
\email{you@math.usf.edu}





\begin{abstract}
A broad set of sufficient conditions consisting of systems of
linear partial differential equations is presented which
guarantees that the Wronskian determinant solves the Korteweg-de
Vries equation in the bilinear form. A systematical analysis is
made for solving the resultant linear systems of second-order and
third-order partial differential equations, along with solution
formulas for their representative systems. The key technique is to
apply variation of parameters in solving the involved
non-homogeneous partial differential equations. The obtained
solution formulas provide us with a comprehensive approach to
construct the existing solutions and many new solutions including
rational solutions, solitons, positons, negatons, breathers,
complexitons and interaction solutions of the Korteweg-de Vries
equation.

\end{abstract}

\maketitle

\def \be {\begin{equation}}
\def \ee {\end{equation}}
\def \bea {\begin{eqnarray}}
\def \eea {\end{eqnarray}}
\def \ba {\begin{array}}
\def \ea {\end{array}}
\def \si {\sigma}
\def \al {\alpha}
\def \la {\lambda}
\def \D {\displaystyle }
\newcommand{\R}{\mathbb{R}}

\section{Introduction}
\setcounter{equation}{0}

Among integrable equations is the celebrated Korteweg-de Vries
(KdV) equation, which serves as a model equation governing weakly
nonlinear long waves whose phase speed attains a simple maximum
for waves of infinite length \cite{DrazinJ-book1989}. It motivates
us to explore beauty hidden in nonlinear differential (and
difference) equations. The remarkable and exceptional discovery of
the inverse scattering transform \cite{AblowitzS-book1981} is one
of important developments in the field of applied mathematics,
which comes from the study of the KdV equation. There are various
algebraic and geometric characteristics that the KdV equation
possesses, for example, infinitely many symmetries and infinitely
many conserved densities \cite{MiuraGK-JMP1968}, the Lax
representation \cite{Lax-CPAM1968}, bi-Hamiltonian structure
\cite{Magri-JMP1978}, loop group \cite{SegalW-IHESPM1985}, and the
Darboux-B\"acklund transformation \cite{MatveevS-book1991}. More
significantly, many physically important solutions to the KdV
equation can be presented explicitly through a simple, specific
form, called the Hirota bilinear form. Such exact solutions
contain solitons \cite{Hirota-PRL1971,Satsuma-JPSJ1979}, rational
solutions \cite{AblowitzS-JMP1978}, positons
\cite{Arkad'evPP-ZNSLOMI1984,Matveev-PLA1992,Kovalyov-NA1988},
negatons \cite{RasinariuSK-JPA1996}, breathers
\cite{Jaworski-PLA1984} and complexitons \cite{Ma-PLA2002}.

Let us consider the KdV equation in its standard form \be
u_{t}-6uu_x+u_{xxx}=0.\label{eq:kdv:TAMS202}\ee Hirota
\cite{Hirota-PRL1971} introduced the transformation (here and in
the rest of the paper, we use the notation $\ln f$ for simplicity,
which does not imply that $f>0$)
\[ u=-2\partial _x^2 \ln f =-2({\ln }f)_{xx}=-\frac
{2(ff_{xx}-f_x^2)}{f^2}, \] and transformed the KdV equation
(\ref{eq:kdv:TAMS202}) into the bilinear form \be
(D_xD_t+D_x^4)f\cdot
f=f_{xt}f-f_tf_x+f_{xxxx}f-4f_{xxx}f_x+3f_{xx}^2=0,
\label{eq:bkdv:TAMS202}\ee which is called the bilinear KdV
equation. Here $D_x$ and $D_t$ are Hirota's bilinear operators
\cite{BulloughC-book1980}, defined by
\[ f(x+h,t+k)g(x-h,t-k)=\sum_{i,j=0}^\infty \frac 1 {i!j!}(D_x^iD_t^j f\cdot g)h^ik^j,\]
which are also closely related to the vertex operators from
elementary particle theory. Strictly speaking, we have \[ \ba {l}
u_{t}-6uu_x+u_{xxx} = -\Bigl[\D \frac {2(D_xD_t+D_x^4)f\cdot
f}{f^2} \Bigr]_x .\ea
 \]
Therefore, if $f$ solves the bilinear KdV equation
(\ref{eq:bkdv:TAMS202}), then $u=-2\partial _x^2 \ln f $ solves
the original KdV equation (\ref{eq:kdv:TAMS202}). The bilinear
form (\ref{eq:bkdv:TAMS202}) looks a little bit more complicated
than the KdV equation (\ref{eq:kdv:TAMS202}) itself, but its
bilinear property, the nearest neighbor of the linear property,
brings us great convenience in constructing explicit exact
solutions. Such a bilinear form exists not only for the KdV
equation, but also for many other physically important soliton
equations, such as the Ablowitz-Kaup-Newell-Segur equations
\cite{AblowitzKNS-SAM1974}, the Kadomtsev-Petviashvili equation
\cite{KadomtsevP-SPD1970}, and the Benney-Roskes equation
\cite{BenneyR-SAM1969} (or the Davey-Stewartson equation
\cite{DaveyS-PRSA1974}).

The Wronskian technique is a powerful tool to construct exact
solutions to bilinear differential (and difference) equations. To
use this technique, we adopt the compact Freeman and Nimmo's
notation \cite{FreemanN-PLA1983}:
 \be
(\widehat{N-1})=(\widehat{N-1};\Phi)=W(\phi
_1,\phi_2,\cdots,\phi_{N})=\left |
\ba {cccc}\phi _1^{(0)}&\phi _1^{(1)}&\cdots &\phi_1^{(N-1)}\\
\phi _2^{(0)}&\phi_2^{(1)}&\cdots
 &\phi_2^{(N-1)}\\ \vdots &\vdots & \ddots & \vdots \\
\phi _{N}^{(0)}&\phi_{N}^{(1)}&\cdots
 &\phi_N^{(N-1)}\ea
\right |,\ N\ge 1,\label{eq:wronskian:TAMS202}\ee where \be
\phi_i^{(0)}=\phi_i,\, \phi _i^{(j)}=\frac {\partial ^j}{\partial
x^j}\phi_i,\ j\ge 1,\,1\le i\le N.\ee Satsuma, Freeman and Nimmo
\cite{Satsuma-JPSJ1979,FreemanN-PLA1983} proved that multi-soliton
solutions to the KdV equation can be expressed through the above
Wronskian determinant, and afterwards, Matveev
\cite{Matveev-PLA1992} generalized the Wronskian determinant which
allows to present another important class of exact solutions,
called positons, to the KdV equation. In using the Wronskian
method to solve the KdV equation, one usually starts from
 \be -\phi_{i,xx}=\lambda _i \phi_i,\ \phi_{i,t}=-4\phi_{i,xxx},
\ 1\le i\le N,\label{eq:basicrequirement:TAMS202} \ee where
$\lambda_i$ are arbitrary real constants, to guarantee that the
Wronskian determinant (\ref{eq:wronskian:TAMS202}) solves the
bilinear KdV equation (\ref{eq:bkdv:TAMS202}) and thus generates a
solution \be u=-2\partial _x ^2 \ln W(\phi
_1,\phi_2,\cdots,\phi_{N})\label{eq:wronskiansol:TAMS202}\ee
 to the KdV
equation (\ref{eq:kdv:TAMS202}). The solution determined by a
Wronskian determinant is called a Wronskian solution. For example,
$f=W(\phi_1,\phi_2,\cdots,\phi_n)$ and $u=-2\partial _x ^2 \ln
W(\phi _1,\phi_2,\cdots,\phi_{N})$ are Wronskian solutions to the
bilinear KdV equation (\ref{eq:bkdv:TAMS202}) and the KdV equation
(\ref{eq:kdv:TAMS202}), respectively, provided that
$\phi_1,\phi_2,\cdots,\phi_n$ satisfy
(\ref{eq:basicrequirement:TAMS202}). Sirianunpiboon et al.
\cite{SirianunpiboonHR-PLA1988} extended the conditions
(\ref{eq:basicrequirement:TAMS202}) to the following \be
-\phi_{i,xx}=\sum_{j=1}^i\lambda _{ij} \phi_j,\
\phi_{i,t}=-4\phi_{i,xxx},\ 1\le i\le N,
\label{eq:lowertriangularcaseofsufficentconditions:TAMS202} \ee
where $\lambda_{ij}$ are arbitrary real constants, in order that
the Wronskian determinant can generate rational function solutions
and their interaction solutions with multi-solitons. The first
group of the above conditions
(\ref{eq:lowertriangularcaseofsufficentconditions:TAMS202}) is a
triangular system of ordinary differential equations. Thus,
solving the system of differential equations
(\ref{eq:lowertriangularcaseofsufficentconditions:TAMS202}) goes
exactly in the same ways for scalar differential equations.

In this paper, we would like to construct explicit exact solutions
to the KdV equation (\ref{eq:kdv:TAMS202}) by its bilinear form
and to illustrate the entire process of construction with a
broader set of sufficient conditions on the Wronskian solutions:
\be -\phi_{i,xx}=\sum_{j=1}^N\lambda _{ij} \phi_j,\
\phi_{i,t}=-4\phi_{i,xxx}+\xi \phi_i, \ 1\le i\le N,
\label{eq:generalcaseofsufficentconditions:TAMS202} \ee where
 $\lambda_{ij}$ and $\xi $ are arbitrary real constants. Here the first group of
conditions is a coupled system of ordinary differential equations
of second-order, and thus needs a more general consideration for
solving the involved differential equations. Note that the
coefficient matrix $\Lambda=(\lambda _{ij})$ in
(\ref{eq:lowertriangularcaseofsufficentconditions:TAMS202}) has
only real eigenvalues, but the coefficient matrix
$\Lambda=(\lambda _{ij})$ in
(\ref{eq:generalcaseofsufficentconditions:TAMS202}) in general has
complex eigenvalues. This is a significant difference between two
sets of conditions, defined by
(\ref{eq:lowertriangularcaseofsufficentconditions:TAMS202}) and
(\ref{eq:generalcaseofsufficentconditions:TAMS202}). The
substantial extension in
(\ref{eq:generalcaseofsufficentconditions:TAMS202}) allows us to
get a much broader class of explicit exact solutions to the KdV
equation (\ref{eq:kdv:TAMS202}). In particular, new exact
solutions called complexitons can be generated from the set of
conditions (\ref{eq:generalcaseofsufficentconditions:TAMS202}),
but not
(\ref{eq:lowertriangularcaseofsufficentconditions:TAMS202}).

Now the whole problem of constructing the Wronskian solutions to
the KdV equation (\ref{eq:kdv:TAMS202}) reduces to solve the
coupled system of partial differential equations
(\ref{eq:generalcaseofsufficentconditions:TAMS202}), which turns
out to be an interesting mathematical problem itself. We shall
analyze solution structures on the system
(\ref{eq:generalcaseofsufficentconditions:TAMS202}) in detail, to
obtain solution formulas for all cases of the system
(\ref{eq:generalcaseofsufficentconditions:TAMS202}). The resultant
fundamental solution formulas provide a direct and comprehensive
approach to construct diverse exact solutions to the KdV equation
(\ref{eq:kdv:TAMS202}), such as rational solutions, solitons,
positons, negatons, breathers, complexitons, and more generally
their interaction solutions.

\section{Sufficient conditions on Wronskian solutions}
\setcounter{equation}{0} \label{sec:kdvsolu:TAMS202}

We begin with stating a broad set of sufficient conditions which
make the Wronskian determinant a solution to the bilinear KdV
equation (\ref{eq:bkdv:TAMS202}).

\begin{theorem}\label{thm:gensolf:TAMS202} Assume that
a group of functions $\phi _i=\phi_i(x,t),\ 1\le i\le N,$
satisfies the following two sets of conditions \bea &&
-\phi_{i,xx}=\sum_{j=1}^N\lambda _{ij}(t)\phi _j ,\ 1\le i\le N,\label{eq:cond1:TAMS202} \\
&& \phi_{i,t}=-4\phi_{i,xxx} +\xi (t)\phi_i,\ 1\le i\le
N,\label{eq:cond2:TAMS202}  \eea simultaneously, where $\lambda
_{ij}(t)$ are arbitrary differentiable real functions of $t$ and $
\xi (t)$ is an arbitrary continuous real function of $t$. Then $f=
(\widehat{N-1})$ defined by (\ref{eq:wronskian:TAMS202}) solves
the bilinear KdV equation (\ref{eq:bkdv:TAMS202}).
\end{theorem}

The conditions (\ref{eq:cond1:TAMS202}) and
(\ref{eq:cond2:TAMS202}) are a generalization to
(\ref{eq:lowertriangularcaseofsufficentconditions:TAMS202})
presented by Sirianunpiboon et al.
\cite{SirianunpiboonHR-PLA1988}. Actually, if $\xi =0$ and the
coefficient matrix $\Lambda =(\lambda _{ij})$ of
(\ref{eq:cond1:TAMS202}) is lower-triangular and independent of
time $t$, then the result of the above theorem boils down to the
result in \cite{SirianunpiboonHR-PLA1988}. The proof of the
theorem needs some basic equalities, and the following notation is
helpful in our deduction and analysis: \bea &&
(\widehat{N-j-1},i_1,\cdots,i_j)=
|\Phi^{(0)},\Phi^{(1)},\cdots,\Phi^{(N-j-1)},\Phi^{(i_1)},\cdots,\Phi^{(i_j)}| \nonumber \\
&&= \det
(\Phi^{(0)},\Phi^{(1)},\cdots,\Phi^{(N-j-1)},\Phi^{(i_1)},\cdots,\Phi^{(i_j)})
,\
 1\le j\le N-1,\eea
where $i_1,\cdots,i_j$ are non-negative integers, and the vectors
of functions $\Phi^{(j)}$ are defined by \be \Phi^{(j)}
=(\phi_1^{(j)},\phi_2^{(j)},\cdots,\phi_N^{(j)})^T,\ 0\le j\le
N-1.\ee We also use the assumption for convenience that if $i<0$,
the column vector $\Phi^{(i)}$ will disappear in the determinant
$\det (\cdots,\Phi^{(i)},\cdots)$.

\begin{lemma}\label{le:basiclemma:TAMS202}
Under the conditions (\ref{eq:cond1:TAMS202}), the following
equalities hold:
\begin{eqnarray}&&
 \sum_{i=1}^N\lambda
 _{ii}(t)(\widehat{N-1})=(\widehat{N-3},N-1,N)-
(\widehat{N-2},N+1),\label{eq:use1} \\&&
 \frac {\partial ^2(\widehat{N-1})}{\partial x^2}=-\sum_{i=1}^N\lambda _{ii}(t)
(\widehat{N-1})+2(\widehat{N-3},N-1,N),\label{eq:use2}\\ &&
\sum_{i=1}^N\lambda
_{ii}(t)(\widehat{N-3},N-1,N)=(\widehat{N-5},N-3,N-2,N-1,N)
\label{eq:use3} \\&& \qquad
+(\widehat{N-3},N,N+1)-(\widehat{N-3},N-1,N+2),\nonumber
\end{eqnarray}
and
\begin{eqnarray}&&
\Bigl(\sum_{i=1}^N\lambda _{ii}(t)\Bigr)^2(\widehat{N-1})
=(\widehat{N-5},N-3,N-2,N-1,N)\label{eq:use4} \\&& \qquad
-(\widehat{N-4},N-2,N-1,N+1)-(\widehat{N-3},N-1,N+2)\nonumber\\&&
\qquad +2(\widehat{N-3},N,N+1)+(\widehat{N-2},N+3).\nonumber
\end{eqnarray}
\end{lemma}
\begin{proof} Note that we have
\[  \sum_{k=1}^N|A|_{kl}=\sum_{k=1}^N|A|^{kl}=\sum_{i,j=1}^N
A_{ij}\frac {\partial ^l a_{ij}}{\partial x^l},
\] where $A=(a_{ij})_{N\times
N},$ and $|A|_{kl}$, $|A|^{kl}$ and $A_{ij}$ denote the
determinant resulting from $|A|$ with its $k$th row differentiated
$l$ times with respect to $x$, the determinant resulting from
$|A|$ with its $k$th column differentiated $l$ times with respect
to $x$, and the co-factor of $a_{ij}$, respectively. Choose $|A|$
as $(\widehat{N-1})$ and $(\widehat{N-3},N-1,N)$, and use the
above equality with $l=2$ and the conditions
(\ref{eq:cond1:TAMS202}). Then we obtain the required equalities
(\ref{eq:use1}) and (\ref{eq:use3}) immediately. A combination of
the equality (\ref{eq:use1}) and the equality
\[ \frac {\partial ^2 (\widehat{N-1})}{\partial x^2}=(\widehat{N-3},N-1,N) +(\widehat{N-2},N+1)\]
leads to the equality
 (\ref{eq:use2}).

We now differentiate (\ref{eq:use1}) twice with respect to $x$ and
utilize (\ref{eq:use2}) and (\ref{eq:use3}). Then a further
computation yields the equality (\ref{eq:use4}). This completes
the proof of the lemma.
\end{proof}

\noindent {\it Proof of Theorem \ref{thm:gensolf:TAMS202}.} By
using the conditions (\ref{eq:cond2:TAMS202}), we obtain that
\bea && f_t=N\xi f-4[(\widehat{N-4},N-2,N-1,N)\nonumber \\
&& \qquad -(\widehat{N-3},N-1,N+1) +(\widehat{N-2},N+2)],\nonumber \\
&&
f_{xt}=N\xi f_x-4[(\widehat{N-5},N-3,N-2,N-1,N)\nonumber \\
&& \qquad -(\widehat{N-3},N,N+1) +(\widehat{N-2},N+3)].\nonumber
\eea Therefore, we can further deduce that
\begin{eqnarray} &&
 \Delta :=
D_x(D_t+D_x^3)f\cdot
f=f_{xt}f-f_tf_x+f_{xxxx}f-4f_{xxx}f_x+3f_{xx}^2\nonumber
\\
&&=[-3(\widehat{N-5},N-3,N-2,N-1,N)+6(\widehat{N-3},N,N+1)-3(\widehat{N-2},N+3)
\nonumber\\
&& \
+3(\widehat{N-4},N-2,N-1,N+1)+3(\widehat{N-3},N-1,N+2)](\widehat{N-1})
\nonumber\\
&&\ -12(\widehat{N-3},N-1,N+1)(\widehat{N-2},N)
\nonumber\\
&&\ +3[(\widehat{N-3},N-1,N)+ (\widehat{N-2},N+1)]^2.\nonumber
\end{eqnarray}
Now using the equalities (\ref{eq:use1}) and (\ref{eq:use4}) in
Lemma \ref{le:basiclemma:TAMS202} and the Laplace expansion of
determinants about the last $N$ rows, we have
\begin{eqnarray}&&
 \Delta = -12(\widehat{N-3},N-1,N+1)
(\widehat{N-2},N)\nonumber\\
&& \quad +12(\widehat{N-3},N,N+1)(\widehat{N-1})
+12(\widehat{N-3},N-1,N)(\widehat{N-2},N+1)\nonumber\\
&&=6\left |\ba {cccccc}\widehat{N-3}&0&N-2&N-1&N&N+1\vspace{2mm}\\
0&\widehat{N-3}&N-2&N-1&N&N+1 \ea \right|=0.\nonumber
\end{eqnarray}
This shows that $f=(\widehat{N-1})$ solves
(\ref{eq:bkdv:TAMS202}). Note that the entries in the last
abbreviated $2N\times 2N$ determinant above are derivatives of
$\Phi$, e.g., $N-1$ denotes $\Phi^{(N-1)}$.
 The
proof is finished.\hfill $\Box $

Theorem \ref{thm:gensolf:TAMS202} tells us that if a group of
functions $\phi_i(x,t),\ 1\le i\le N$, satisfies the conditions
(\ref{eq:cond1:TAMS202}) and (\ref{eq:cond2:TAMS202}), then we can
get a solution $f=W(\phi_1,\phi_2,\cdots,\phi_N)$ to the bilinear
KdV equation (\ref{eq:bkdv:TAMS202}). The conditions
(\ref{eq:cond2:TAMS202}) and (\ref{eq:cond1:TAMS202}) are two
linear systems of second-order and third-order partial
differential equations. We are going to solve these linear systems
explicitly. It is rather difficult to deal with the case where the
coefficient matrix $\Lambda$ of (\ref{eq:cond1:TAMS202}) has
complex eigenvalues. However, when $-\Lambda$ is a positive
definite constant matrix, the situation is easy and the solution
formula of (\ref{eq:cond1:TAMS202}) and (\ref{eq:cond2:TAMS202})
can be expressed as \be \Phi=\textrm{e}^{\int_0^t\xi
(t')\,dt'}\textrm{e}^{\sqrt{-\Lambda}\,x+4\Lambda \sqrt{-\Lambda
}\,t}\Phi_0,\ee  where $\sqrt{-\Lambda}$ is the square root of
$-\Lambda$ and  $\Phi_0$ is an arbitrary initial vector. The
corresponding Wronskian solutions to the KdV equation
(\ref{eq:kdv:TAMS202}), \be u=-2\partial _x^2 \ln f=-2\partial
_x^2 \ln (\widehat {N-1};\Phi), \ee contain soliton and negaton
solutions.

Before we proceed to solve (\ref{eq:cond1:TAMS202}) and
(\ref{eq:cond2:TAMS202}), let us observe the Wronskian
determinants and solutions more carefully.

\noindent {\bf Observation I}: From the compatibility conditions
$\phi_{i,txx}=\phi_{i,xxt},\ 1\le i\le N,$ of the conditions
(\ref{eq:cond1:TAMS202}) and (\ref{eq:cond2:TAMS202}), we have the
equality \be \sum_{j=1}^N \lambda_{ij,t}\phi_j=0,\ 1\le i\le N,
\ee
 and thus it is easy to see that the Wronskian
determinant $W(\phi_1,\phi_2,\cdots,\phi_N)$ becomes zero if there
is at least one entry $\lambda _{ij}$ satisfying $\lambda
_{ij,t}\ne 0$.

\noindent {\bf Observation II}: If we make use of the
transformation \be  \tilde {\phi}_i=\textrm{e}^{-\int_0^t\xi
(s)\,ds} {\phi}_i,\ 1\le i\le N,\ee  the differential equations in
(\ref{eq:cond2:TAMS202}) can be put into
\[
{\tilde {\phi}}_{i,t}=-4 {\tilde {\phi}}_{i,xxx} ,\ 1\le i\le N,
\]
but the differential equations (\ref{eq:cond1:TAMS202}) do not
change. Obviously, the resultant Wronskian solutions to the KdV
equation are the same: \be u=-2\partial _x ^2 \ln
W(\phi_1,\phi_2,\cdots,\phi_N) = \tilde {u}=-2\partial _x ^2  \ln
W(\tilde{\phi}_1,\tilde{\phi}_2,\cdots,\tilde{\phi}_N).\ee But the
Wronskian determinants $W(\phi_1,\phi_2,\cdots ,\phi_N)$ and
$W(\tilde{\phi}_1,\tilde{\phi}_2,\cdots,\tilde{\phi}_N)$ are
different, and thus they gives rise to different solutions to the
bilinear KdV equation (\ref{eq:bkdv:TAMS202}).

\noindent {\bf Observation III}:  If the coefficient matrix
$\Lambda=(\lambda _{ij})$ is similar to another matrix
$M=(\mu_{ij})$ under an invertible constant matrix $P$, let us say
$\Lambda =P^{-1}M P$, then $\tilde {\Phi}=P\Phi$ solves \be \tilde
{\Phi}_{xx}=M \tilde {\Phi},\ \tilde {\Phi}_t=-4\tilde
{\Phi}_{xxx}+\xi \tilde {\Phi},\
 \ee
and the resultant Wronskian solutions to the KdV equation are also
the same: \begin{eqnarray} && u(\Lambda )=-2\partial _x ^2\ln
|\Phi^{(0)},\Phi^{(1)},\cdots,\Phi^{(N-1)} |
\\ &&
\qquad\  =-2\partial _x ^2 \ln
|P\Phi^{(0)},P\Phi^{(1)},\cdots,P\Phi^{(N-1)}|=u(M).\nonumber
\end{eqnarray}

It follows from Observations II and III that different Wronskian
determinants may lead to the same solution to the KdV equation.
Therefore, based on Observation I, in order to construct Wronskian
solutions to the KdV equation by its bilinear form, we only need
to consider the reduced case of (\ref{eq:cond2:TAMS202}) and
(\ref{eq:cond1:TAMS202}) under $\xi =0$ and $d\Lambda /dt=0$,
i.e., the following conditions
\begin{equation}
 -\phi_{i,xx}=\sum_{j=1}^N\lambda  _{ij}\phi _j
,\ \phi_{i,t}=-4\phi_{i,xxx},\ 1\le i\le
N,\label{eq:sscofbkdv:TAMS202}
\end{equation}
where $\lambda _{ij}$ are arbitrary real constants. On the other
hand, the Jordan form of a real matrix has the following two types
of blocks: \bea && \left[\ba {cccc}\lambda_i & & & 0
\vspace{2mm}\\
1 & \lambda_i  &  &  \vspace{2mm}\\
&\ddots& \ddots&  \vspace{2mm}\\
0 & & 1& \lambda_i  \ea \right]_{k_i\times k_i},
\label{eq:1stJordanblock:TAMS202}
\\
&& \left[\ba {cccc} A_i & & & 0
\vspace{2mm}\\
 I_2&A_i& & \vspace{2mm}\\
& \ddots& \ddots& \vspace{2mm}\\
0 &&I_2 &A_i \ea \right]_{l_i\times l_i},\ A_i=\left[\ba
{cc}\alpha _i&-\beta _i\vspace{2mm}\\ \beta _i&\alpha _i \ea
\right],\ I_2=\left[\ba {cc}1&0\vspace{2mm}\\
0&1 \ea \right],
 \label{eq:2rdJordanblock:TAMS202} \eea where
$\lambda_i$,  $\alpha_i $ and $\beta_i >0$ are all real constants.
The first type of blocks has the real eigenvalue $\lambda_i$ with
algebraic multiplicity $k_i$, and the second type of blocks has
the complex eigenvalues $\lambda _{i}^{\pm} =\alpha_i\pm \beta _i
\sqrt{-1}\,$ with algebraic multiplicity $l_i$. Note that an
eigenvalue of the coefficient matrix $\Lambda =(\lambda _{ij})$ is
also an eigenvalue of the Schr\"odinger operator $-\frac {\partial
^2}{\partial x^2}+u$ with zero potential. We will present solution
formulas for the system of differential equations defined by
(\ref{eq:sscofbkdv:TAMS202}), according to the situations of
eigenvalues of the coefficient matrix. Now, based on Observation
III, all we need to do is to solve a group of subsystems of the
sufficient conditions on the Wronskian solutions, whose
coefficient matrices are of the forms
(\ref{eq:1stJordanblock:TAMS202}) and
(\ref{eq:2rdJordanblock:TAMS202}).

Let us summarize the above analysis as the following theorem.
\begin{theorem} For the KdV equation (\ref{eq:kdv:TAMS202}),
all Wronskian solutions generated from the conditions
(\ref{eq:cond1:TAMS202}) and (\ref{eq:cond2:TAMS202}) are among
the Wronskian solutions associated with the special cases of the
conditions (\ref{eq:sscofbkdv:TAMS202}) whose coefficient matrices
$\Lambda =(\lambda _{ij})$ are of the Jordan form consisting of
two types of Jordan blocks in (\ref{eq:1stJordanblock:TAMS202})
and (\ref{eq:2rdJordanblock:TAMS202}).
\end{theorem}

In the next section, we shall consider how to solve the two types
of subsystems associated with the Jordan blocks in
(\ref{eq:1stJordanblock:TAMS202}) and
(\ref{eq:2rdJordanblock:TAMS202}), and present solution formulas
for their representative systems generating Wronskian solutions to
the KdV equation (\ref{eq:kdv:TAMS202}).

\section{Solution formulas for the representative systems}
\setcounter{equation}{0}

In this section, we would like to consider the construction of
solutions to the associated system of differential equations
defined by (\ref{eq:sscofbkdv:TAMS202}). Based on the form of two
types of Jordan blocks in (\ref{eq:1stJordanblock:TAMS202}) and
(\ref{eq:2rdJordanblock:TAMS202}), a basic idea to solve the
system (\ref{eq:sscofbkdv:TAMS202}) is to use a recursion process.
That is to solve (\ref{eq:sscofbkdv:TAMS202}) from $\phi_1$ to
$\phi_N$, one by one when the coefficient matrix $\Lambda
=(\lambda _{ij})$ has real eigenvalues or pair by pair when the
coefficient matrix $\Lambda =(\lambda _{ij})$ has complex
eigenvalues. Therefore, the entire problem is divided into two
subproblems - to solve the following two representative systems of
non-homogeneous differential equations: \be -\phi_{xx}=\lambda
\phi +f,\ \phi_{t}=-4\phi_{xxx},\
\label{eq:1stsubproblem:TAMS202}\ee and \be \left \{\ba {l}
 -\phi_{1,xx}=\alpha \phi_1-\beta\phi_2+f_1,\vspace{2mm} \\
 -\phi_{2,xx}=\beta \phi_1+\alpha \phi_2+f_2,\vspace{2mm}\\
\phi_{i,t}=-4\phi_{i,xxx},\ i=1,2,
 \ea
 \right.
\label{eq:2ndsubproblem:TAMS202} \ee where $\lambda$, $\alpha$ and
$\beta
>0$ are real constants, and $f=f(x,t)$, $f_1=f_1(x,t)$ and
$f_2=f_2(x,t)$ are three given functions satisfying the
compatibility condition $g_t=-4g_{xxx}$. The whole system
(\ref{eq:sscofbkdv:TAMS202}) whose coefficient matrix is of Jordan
form decouples into a group of systems determined by
(\ref{eq:1stsubproblem:TAMS202}) and
(\ref{eq:2ndsubproblem:TAMS202}).

\subsection{The case of real eigenvalues}

First, let us consider the first representative system
(\ref{eq:1stsubproblem:TAMS202}). In terms of the eigenvalue
$\lambda $, we will establish solution formulas for three
situations of the representative system
(\ref{eq:1stsubproblem:TAMS202}).

{\bf Zero eigenvalue:} The representative system
(\ref{eq:1stsubproblem:TAMS202}) in this case is  \be
 \phi_{xx}=f,\ \phi_{t}=-4\phi_{xxx}, \label{eq:firstcase} \ee where
$f=f(x,t)$ is a given function satisfying the compatibility
condition $f_t=-4f_{xxx}$. Immediately from the first equation in
(\ref{eq:firstcase}), we can get \be
\phi=\int_0^x\int_0^{x'}f(x'',t)dx''dx'
+c(t)x+d(t).\label{eq:firstphi} \ee Note that $f_t=-4f_{xxx}$, and
thus the second equation of (\ref{eq:firstcase}) equivalently
requires
\[
c_{t}=-4f_{xx}|_{x=0},\ d_{t}=-4f_{x}|_{x=0}.
\]
Then, we obtain \be c(t)=-4\int_0^tf_{xx}(0,t')\,dt' +c_0,
 \
d(t)=-4\int_0^tf_{x}(0,t')\,dt' +d_0,\label{eq:firstcandd} \ee
where $c_{0}$ and $d_{0}$ are arbitrary real constants. Summing
up, we have the following theorem.

\begin{theorem}Let $f=f(x,t)$ be
a given function satisfying the compatibility condition
$f_t=-4f_{xxx}$. Then the system of differential equations
(\ref{eq:firstcase}) has the general solution given by
(\ref{eq:firstphi}) and (\ref{eq:firstcandd}).
\end{theorem}

{\bf Negative eigenvalue:} In this case, the representative system
(\ref{eq:1stsubproblem:TAMS202}) reads as
 \be \phi_{xx}=\al \phi+
f,\ \phi_{t}=-4\phi_{xxx},\ \alpha
>0,\label{eq:secondcase} \ee where $\al $ is a constant
and $f$ is a given function satisfying the compatibility condition
$f_t=-4f_{xxx}$. From two basic solutions
$\textrm{e}^{\sqrt{\alpha}x}$ and $\textrm{e}^{-\sqrt{\alpha}x}$
of $\phi_{xx}=\alpha \phi$, we get a general solution of the first
differential equation of (\ref{eq:secondcase}):
 \begin{eqnarray}\label{eq:secondphi}
\phi&=&\Bigl[\frac 1{2\sqrt{\al }} \int_0^xf(x',t)
\textrm{e}^{-\sqrt{\al }\,x'}dx'+c(t)\Bigr]\textrm{e}^{\sqrt{\al
}\,x}
\\ &&
 -\Bigl[\frac 1{2\sqrt{\al }}
\int_0^xf(x',t) \textrm{e}^{\sqrt{\al
}\,x'}dx'+d(t)\Bigr]\textrm{e}^{-\sqrt{\al }\,x}, \nonumber
\end{eqnarray} by variation of parameters.
 And then
noting that $f_t=-4f_{xxx}$, by a direct but lengthy computation,
we find that the second differential equation of
(\ref{eq:secondcase}) equivalently requires \bea && c_{t}=-4\al
\sqrt{\al }\,c-2(
\frac 1 {\sqrt{\al }}\,f_{xx}+f_{x}+\sqrt{\al }\,f)|_{x=0},\nonumber \\
&& d_{t}=4\al \sqrt{\al }\,d+2(- \frac 1 {\sqrt{\al
}}\,f_{xx}+f_{x}-\sqrt{\al }\,f)|_{x=0}.\nonumber \eea These are
two linear systems for $c$ and $d$, respectively. Hence, we can
immediately obtain the solution formulas for $c$ and $d$: \bea &&
c(t)=\textrm{e}^{-4\alpha \sqrt{\alpha }\, t}
 \Bigl[c_0-2\int_0^t\textrm{e}^{4\alpha \sqrt{\alpha }\, t'}
(\frac 1 {\sqrt{\al }}\,f_{xx}+f_{x}+\sqrt{\al }\,f)(0,t'))\,dt'
\Bigr] ,\qquad \label{eq:secondc}
\\ && d(t)=
\textrm{e}^{4\alpha \sqrt{\alpha }\, t}
 \Bigl[d_0+2\int_0^t\textrm{e}^{-4\alpha \sqrt{\alpha }\, t'}
(-\frac 1 {\sqrt{\al }}\,f_{xx}+f_{x}-\sqrt{\al }\,f)(0,t'))\,dt'
\Bigr],\label{eq:secondd} \eea where $c_0$ and $d_0$ are arbitrary
real constants. We conclude this result as follows.

\begin{theorem}
\label{thm:seofsecondsituation} Let $f=f(x,t)$ be a given function
satisfying the compatibility condition $ f_t=-4f_{xxx}.$ Then the
system of differential equations (\ref{eq:secondcase}) has the
general solution given by (\ref{eq:secondphi}), (\ref{eq:secondc})
and (\ref{eq:secondd}).
\end{theorem}

{\bf Positive eigenvalue:} The representative system
(\ref{eq:1stsubproblem:TAMS202}) in this case is  \bea
&&\phi_{xx}=-\alpha \phi +f,  \ \alpha >0, \label{eq:thirdcase1:TAMS202}\\
&&\phi_{t}=-4\phi_{xxx}, \label{eq:thirdcase2:TAMS202} \eea where
$\alpha $ is a constant and $f=f(x,t)$ is a given function
satisfying the compatibility condition $ f_t=-4f_{xxx}.$ To solve
the differential equation (\ref{eq:thirdcase1:TAMS202}), we can
start from two basic solutions $\sin \sqrt{\alpha }x$ and $\cos
\sqrt{\alpha }x$ of the corresponding homogeneous equation of
(\ref{eq:thirdcase1:TAMS202}), and then use the method of
variation of parameters. A general solution is listed in the
following lemma for reference.
\begin{lemma}
\label{le:negativeeigenvaluesecondODE:TAMS202} For any given
function $g=g(x)$ and any positive constant $\alpha $, the
second-order ordinary differential equation $\psi_{xx}=-\alpha
\psi +g$ has the general solution \[ \psi=\int_0^x \frac {\sin
\sqrt{\alpha}(x-x')} {\sqrt{\alpha}} g(x')\, dx' + c\cos
\sqrt{\alpha}x +d \sin \sqrt{\alpha} x ,
\label{eq:negativeeigenvaluesecondODE} \] where $c$ and $d$ are
arbitrary real constants.
\end{lemma}

Now to solve the partial differential equation
(\ref{eq:thirdcase2:TAMS202}), we start with the solution of
(\ref{eq:thirdcase1:TAMS202}) given by Lemma
\ref{le:negativeeigenvaluesecondODE:TAMS202} and use the method of
variation of parameters again. Therefore, the solution of
(\ref{eq:thirdcase2:TAMS202}) is assumed to be
 \be \phi=\int_0^x
\frac {\sin \sqrt{\alpha}(x-x')} {\sqrt{\alpha}}f(x',t)\, dx' +
c(t)\cos \sqrt{\alpha}x +d(t) \sin \sqrt{\alpha}x .
\label{eq:solphiofnegativesituation:TAMS202} \ee Directly we can
compute that \bea \phi_{xxx}&=& f_x-\int_0^x f(x',t) (\alpha \cos
\sqrt{\alpha}(x-x')) \, dx' \nonumber \\ && + c(t) {\alpha
\sqrt{\alpha }} \sin \sqrt{\alpha}x-
d(t) {\alpha \sqrt{\alpha }}\cos \sqrt{\alpha}x ,\nonumber \\
\phi_t& =&4f_{xx}|_{x=0} \frac {\sin
\sqrt{\alpha}x}{\sqrt{\alpha}} +4f_x|_{x=0}\cos \sqrt{\alpha}x -
4f|_{x=0}(\sqrt{\alpha} \sin \sqrt{\alpha}x) \nonumber \\
&& -4f_x+\int_0^x  4f(x',t) (\alpha\cos \sqrt{\alpha}(x-x'))\, dx'
\nonumber \\
&& + c_t\cos \sqrt{\alpha}x +d_t\sin \sqrt{\alpha}x ,
 \nonumber \eea
where $f_t=-4f_{xxx}$ has been used for the computation of
$\phi_t$. Therefore, the differential equation
(\ref{eq:thirdcase2:TAMS202}) equivalently requires
\begin{eqnarray} &&
 \frac 4 {\sqrt{\alpha}}f_{xx}|_{x=0}-4\sqrt{\alpha}f|_{x=0}+d_t=-4 \alpha \sqrt{\alpha} \,
 c,\\
 &&
 4f_x|_{x=0}+c_t=4\alpha \sqrt{\alpha}\, d. \label{eq:eqofd:TAMS202}
\end{eqnarray}
 Then it follows that \[ c_{tt}+16 \alpha ^3c+ h=0,\]
 where the function
 $h=h(t)$ is defined by \be  h:=
16 \alpha f_{xx}|_{x=0}-16 \alpha ^2
f|_{x=0}+4f_{xt}|_{x=0}.\label{eq:defofgin2rdODEofc:TAMS202} \ee
Again by Lemma \ref{le:negativeeigenvaluesecondODE:TAMS202}, we
obtain
\[
c(t)= -\int_0^t \frac {\sin 4 \alpha \sqrt{\alpha}(t-t')}{4\alpha
\sqrt{\alpha}} h(t')\, dt' +c_0\cos 4 \alpha
\sqrt{\alpha}t+d_0\sin 4 \alpha \sqrt{\alpha}t,
\]
where $c_0$ and $d_0$ are arbitrary real constants. Noting that
$h$ is given by (\ref{eq:defofgin2rdODEofc:TAMS202}), we have \bea
c(t) &=& f_x(0,0)\frac {\sin 4 \alpha \sqrt{\alpha}t}{\alpha
\sqrt{\alpha}} -4\int_0^t
f_x(0,t') \cos 4 {\alpha \sqrt{\alpha}}(t-t') \, dt' \label{eq:c(t)ofthirdcase:TAMS202} \\
&& - 4\int _0^t (\frac 1
{\sqrt{\alpha}}f_{xx}-\sqrt{\alpha}f)(0,t')
 \sin 4 \alpha \sqrt{\alpha}(t-t') \, dt'
\nonumber \\
&& +c_0\cos 4 \alpha \sqrt{\alpha}t+d_0\sin 4 \alpha
\sqrt{\alpha}t, \nonumber \eea and then by
(\ref{eq:eqofd:TAMS202}), we have \bea  d(t)&=& f_x(0,0)\frac
{\cos 4 \alpha \sqrt{\alpha}t}{\alpha \sqrt{\alpha}} +4\int_0^t
f_x(0,t') \sin 4 {\alpha \sqrt{\alpha}}(t-t') \, dt'  \label{eq:d(t)ofthirdcase:TAMS202}\\
&& -4\int _0^t (\frac 1
{\sqrt{\alpha}}f_{xx}-\sqrt{\alpha}f)(0,t')
 \cos 4 \alpha \sqrt{\alpha}(t-t') \, dt'
\nonumber \\
&& -c_0\sin 4 \alpha \sqrt{\alpha}t+d_0\cos 4 \alpha
\sqrt{\alpha}t, \nonumber \eea where $c_0$ and $d_0$ are
independent of $x$ and $t$.

Finally summing up, we have the following theorem on the general
solution of the system of differential equations
(\ref{eq:thirdcase1:TAMS202}) and (\ref{eq:thirdcase2:TAMS202}).

\begin{theorem} Let $f=f(x,t)$ be a given function satisfying the compatibility condition
$f_t=-4f_{xxx}$, and $\alpha $ be a positive constant. Then the
system of differential equations (\ref{eq:thirdcase2:TAMS202}) and
(\ref{eq:thirdcase1:TAMS202}) has the general solution given by
(\ref{eq:solphiofnegativesituation:TAMS202}) with $c(t)$ and
$d(t)$ being shown by (\ref{eq:c(t)ofthirdcase:TAMS202}) and
(\ref{eq:d(t)ofthirdcase:TAMS202}), respectively.
\end{theorem}

\subsection{The case of complex eigenvalues}

Let us now consider the second representative system
(\ref{eq:2ndsubproblem:TAMS202}). To present its solution formula,
we first solve the following coupled system of second-order,
non-homogeneous ordinary differential equations \be \left \{ \ba
{l}
-\phi_{1,xx}=\alpha \phi_1-\beta \phi _2 + f_1,\vspace{2mm} \\
-\phi_{2,xx}=\beta \phi_1+\alpha \phi _2 + f_2, \ea \right.
\label{eq:special2ndorderODEs}\ee where $\alpha $ and $\beta
> 0$ are real constants, and $f_1$ and $f_2$ are two given functions of $x$.
Since $\beta >0$, its coefficient matrix
\[A=\left[\ba {cc} \alpha & -\beta \vspace{2mm}\\ \beta & \alpha \ea \right]\]
has two complex eigenvalues $ \la ^{\pm}=\alpha \pm \beta i $.

\begin{theorem}
The coupled non-homogenous system of second-order ordinary
differential equations (\ref{eq:special2ndorderODEs}) has the
general solution \begin{eqnarray} &&
\phi_1=\phi_{1}^h+\phi_{1}^s,\ \phi_2=\phi_{2}^h+\phi_{2}^s,
\label{eq:0compofsolofspecial2ndorderODEs:TAMS202}\\ &&
 \phi_{1}^h=(D_1\cos \delta x +D_2\sin \delta
x)\textrm{e}^{\Delta x}+ (D_3\cos \delta x +D_4\sin \delta
x)\textrm{e}^{-\Delta
x},\label{eq:2compofsolofspecial2ndorderODEs:TAMS202}
\\ &&
\phi_{2}^h=(-D_2\cos \delta x +D_1\sin \delta x)\textrm{e}^{\Delta
x}+ (-D_4\cos \delta x +D_3\sin \delta x)\textrm{e}^{-\Delta x},
\label{eq:4compofsolofspecial2ndorderODEs}
\\ &&
 \phi_{1}^s=-\frac 1 {\Delta }\int_0^x\bigl[f_1(y)\cos\delta(x-y)-f_2(y)
 \sin\delta(x-y)\bigr] \sinh \Delta (x-y)
  \,dy
,\label{eq:1compofsolofspecial2ndorderODEs:TAMS202}
\\ &&
\phi_{2}^s=-\frac 1 {\Delta }\int_0^x\bigl[f_1(y)\sin
\delta(x-y)+f_2(y)
 \cos\delta(x-y)\bigr] \sinh \Delta (x-y)
  \,dy
,\label{eq:3compofsolofspecial2ndorderODEs:TAMS202} \end{eqnarray}
where two constants $\Delta $ and $\delta $ are given by
 \be \Delta =\sqrt{\frac {\sqrt{\alpha ^2+\beta
^2}-\alpha }{2}},\ \delta =\sqrt{\frac {\sqrt{\alpha ^2+\beta
^2}+\alpha }{2}}, \label{eq:defofDeltaanddelta}\ee and $D_i,\,1\le
i\le 4$, are arbitrary real constants.
\end{theorem}

\begin{proof} First let us consider the homogenous case of
(\ref{eq:special2ndorderODEs}), i.e., the case of
(\ref{eq:special2ndorderODEs}) with $f_1=f_2=0$. Therefore, we
have
 \be  \phi_2=
\frac 1 \beta \phi_{1,xx}+\frac \alpha \beta \phi_1.\ee  Then
replacing $\phi_2$ with
this expression in
 the second equation of the homogenous case leads to
the following fourth-order differential equation
 \be \phi_{1,xxxx}+2\alpha
\phi_{1,xx}+(\alpha ^2+\beta ^2)\phi_1 =0.
\label{eq:4thorderlinearODE}
 \ee
Its characteristic polynomial \[ p(\la )=\la ^4+ 2\alpha \la ^2+
(\alpha ^2+\beta ^2)=0\] has four complex roots
\[
\la _1^{\pm }=\Delta \pm \delta \sqrt{-1},\ \la _{2}^{\pm
}=-\Delta \pm \delta \sqrt{-1},
\]
where $\Delta $ and $\delta $ are defined by
(\ref{eq:defofDeltaanddelta}). Therefore, we have the general
solution $\phi_1^h$ and $\phi_2^h$, as given in the theorem, for
the homogeneous case of (\ref{eq:special2ndorderODEs}).

To construct a special solution, $\phi_1^s$ and $\phi_2^s$, for
the non-homogenous case of (\ref{eq:special2ndorderODEs}), we
adopt the method of variation of parameters for $D_i$, $1\le i\le
4$, and thus assume that \be \left\{ \begin{array}{l}
\phi_{1}^s=(D_1\cos \delta x +D_2\sin \delta x)\textrm{e}^{\Delta
x}+ (D_3\cos \delta x +D_4\sin \delta x)\textrm{e}^{-\Delta x},
\vspace{2mm}\\
\phi_{2}^s=(-D_2\cos \delta x +D_1\sin \delta x)\textrm{e}^{\Delta
x}+ (-D_4\cos \delta x +D_3\sin \delta x)\textrm{e}^{-\Delta x},
 \end{array}\right.
 \label{eq:secialsolforphi1andphi2:TAMS202}
\ee where $D_i$, $1\le i \le 4,$  are viewed as functions of $x$.
Then we can make the following computation
 \bea
 \phi_{1,x}^s&=&\bigl[(\Delta D_1+\delta D_2)\cos \delta x +(
\Delta D_2-\delta D_1)\sin \delta x\bigr]\textrm{e}^{\Delta x}
\label{eq:phi1_x:TAMS202}
\\ && + (\delta D_4-\Delta D_3)\cos \delta x -(\delta D_3+\Delta
D_4)\sin \delta x\bigr]
\textrm{e}^{-\Delta x},\nonumber \\
\phi_{1,xx}^s&=&  \bigl[\Delta (\Delta D_1+\delta
D_2)+\delta(\Delta D_2-\delta D_1)\bigr] (\cos \delta x
)\textrm{e}^{\Delta x} \label{eq:phi1_xx:TAMS202}
 \\ && +\bigl[
\Delta (\Delta D_2-\delta D_1)-\delta (\Delta D_1+\delta
D_2)\bigr]
(\sin \delta x)\textrm{e}^{\Delta x} \nonumber \\
&& + \bigl[-\delta (\delta D_3+\Delta D_4)-\Delta (\delta
D_4-\Delta D_3)\bigr](\cos \delta x ) \textrm{e}^{-\Delta x}
\nonumber \\ && +\bigl[-\delta (\delta D_4-\Delta D_3)+
\Delta (\delta D_3+\Delta D_4)\bigr](\sin \delta x )\textrm{e}^{-\Delta x}\nonumber \\
&=&(\alpha D_1+\beta D_2)(\cos \delta x)\textrm{e}^{\Delta x} +
(-\beta D_1+\alpha D_2)(\sin \delta x)\textrm{e}^{\Delta
x}\nonumber\\
&&+(\alpha D_3-\beta D_4)(\cos \delta x)\textrm{e}^{-\Delta x}+
(\beta D_3+\alpha D_4)(\sin \delta x)\textrm{e}^{-\Delta x}
\nonumber \\
&:= & (\tilde {D}_{1}\cos \delta x +\tilde {D}_{2}\sin \delta
x)\textrm{e}^{\Delta x}+ (\tilde {D}_3\cos \delta x +\tilde
{D}_4\sin \delta x)\textrm{e}^{-\Delta x},\nonumber
\\
\phi_{2,x}^s&=&\bigl[(-\Delta D_2+\delta D_1)\cos \delta x +(
\Delta D_1+\delta D_2)\sin \delta x\bigr]\textrm{e}^{\Delta x}
\\ && + (\delta D_3+\Delta D_4)\cos \delta x +(\delta D_4-\Delta
D_3)\sin \delta x\bigr]
\textrm{e}^{-\Delta x},\nonumber
\\
\quad \phi_{2,xx}^s &=&(-\alpha D_2+\beta D_1)(\cos \delta
x)\textrm{e}^{\Delta x} + (\beta D_2+\alpha D_1)(\sin \delta
x)\textrm{e}^{\Delta x}
\\
&&-(\alpha D_4+\beta D_3)(\cos \delta x)\textrm{e}^{-\Delta x}-
(\beta D_4-\alpha D_3)(\sin \delta x)\textrm{e}^{-\Delta
x}.\nonumber
 \eea In the above computation, the method of
variation of parameters required that \bea
 && (D_{1,x}\cos \delta x +D_{2,x}\sin \delta x)\textrm{e}^{\Delta x}+
(D_{3,x}\cos \delta x +D_{4,x}\sin \delta x)\textrm{e}^{-\Delta x}=0,\nonumber\\
&& \bigl[(\Delta D_{1,x}+\delta D_{2,x})\cos \delta x +( \Delta
D_{2,x}-\delta D_{1,x})\sin \delta x\bigr]\textrm{e}^{\Delta x}
\nonumber \\ && + (\delta D_{4,x}-\Delta D_{3,x})\cos \delta x
-(\delta D_{3,x}+\Delta D_{4,x}) \sin \delta x\bigr]
\textrm{e}^{-\Delta x}=-f_1,\nonumber
\\
&& (-D_{2,x}\cos \delta x +D_{1,x}\sin \delta x)\textrm{e}^{\Delta
x}-
(D_{4,x}\cos \delta x -D_{3,x}\sin \delta x)\textrm{e}^{-\Delta x}=0,\ \ \quad \nonumber\\
&& \bigl[(-\Delta D_{2,x}+\delta D_{1,x})\cos \delta x +( \Delta
D_{1,x}+\delta D_{2,x})\sin \delta x\bigr]\textrm{e}^{\Delta x}
\nonumber \\ && + (\delta D_{3,x}+\Delta D_{4,x})\cos \delta x
+(\delta D_{4,x}-\Delta D_{3,x}) \sin \delta x\bigr]
\textrm{e}^{-\Delta x}=-f_2. \nonumber \eea Solving this linear
system for $D_{i,x}$ leads to the expressions:
\[
\ba {l} D_{1,x}= -\D \frac 1{2\Delta }\bigl(f_1 \cos \delta x+ f_2\sin
\delta x \bigr)\textrm{e}^{-\Delta x} , \vspace{2mm}\\
D_{2,x}= -\D \frac 1{2\Delta }\bigl(f_1\sin \delta x-f_2\cos \delta x\bigr)
\textrm{e}^{-\Delta x} , \vspace{2mm}\\
D_{3,x}= \D \frac 1{2\Delta }\bigl(f_1\cos \delta x+ f_2\sin \delta x \bigr)
\textrm{e}^{-\Delta x}=-D_{1,x} , \vspace{2mm}\\
D_{4,x}=\D \frac 1{2\Delta }\bigl(f_1 \sin \delta x -f_2 \cos
\delta x \bigr)\textrm{e}^{-\Delta x} =-D_{2,x}.\ea \] Then
integrating them to obtain the expressions for $D_{i}$ and
inserting the resultant expressions of $D_i$ into
(\ref{eq:secialsolforphi1andphi2:TAMS202}) yield the special
solution, $\phi_1^s$ and $\phi_2^s$, given in the theorem. The
proof is finished.
\end{proof}

Let us then present the solution formula for the second
representative system (\ref{eq:2ndsubproblem:TAMS202}), which can
be used to generate new type Wronskian solutions to the KdV
equation.

\begin{theorem}\label{thm:expressionofphi_1andphi_2ofcaseofcomplexeigenvalues:TAMS202}
Assume that $f_1=f_1(x,t)$ and $f_2=f_2(x,t)$ are two given
functions satisfying the compatibility conditions
$f_{1,t}=-4f_{1,xxx}$ and $ f_{2,t}=-4f_{2,xxx}$. Then the general
solution to the system of differential equations
(\ref{eq:2ndsubproblem:TAMS202})
 is
given by
(\ref{eq:0compofsolofspecial2ndorderODEs:TAMS202})-(\ref{eq:3compofsolofspecial2ndorderODEs:TAMS202})
 with the coefficients $D_i=D_i(t)$, $1\le i\le 4$, being determined by
 \bea && \left[ \ba {c} D_1\vspace{2mm} \\ D_2\ea \right]=
\textrm{e}^{\tilde{\alpha }t}
\left[ \ba {cc} \cos \tilde{\beta  }t & -\sin \tilde{\beta  }t  \vspace{2mm} \\
\sin \tilde{\beta  }s & \cos \tilde{\beta  }t \ea \right] \left[
\ba {c} D_{10}\vspace{2mm} \\ D_{20}\ea \right]
 \label{eq:1solofD_i1to4inthecaseoftildebetanot=0:TAMS202} \\
&&\qquad +\int_0^t \textrm{e}^{\tilde{\alpha }(t-s)}
\left[ \ba {cc} \cos \tilde{\beta  }(t-s) & -\sin \tilde{\beta  }(t-s)  \vspace{2mm} \\
\sin \tilde{\beta  }(t-s) & \cos \tilde{\beta  }(t-s) \ea \right]
\left[ \ba {c} p_1(s)\vspace{2mm} \\ p_{2}(s)\ea \right]\,
ds,\nonumber
\\ &&
\left[ \ba {c} D_3\vspace{2mm} \\ D_4\ea \right]=
\textrm{e}^{-\tilde{\alpha }t}
\left[ \ba {cc} \cos \tilde{\beta  }t & -\sin \tilde{\beta  }t  \vspace{2mm} \\
\sin \tilde{\beta  }s & \cos \tilde{\beta  }t \ea \right] \left[
\ba {c} D_{30}\vspace{2mm} \\ D_{40}\ea \right]  \label{eq:2solofD_i1to4inthecaseoftildebetanot=0:TAMS202}\\
&&\qquad +\int_0^t \textrm{e}^{-\tilde{\alpha }(t-s)}
\left[ \ba {cc} \cos \tilde{\beta  }(t-s) & -\sin \tilde{\beta  }(t-s)  \vspace{2mm} \\
\sin \tilde{\beta  }(t-s) & \cos \tilde{\beta  }(t-s) \ea \right]
\left[ \ba {c} p_3(s)\vspace{2mm} \\ p_{4}(s)\ea \right]\, ds,
\nonumber
 \eea
where $D_{i0},\,1\le i\le 4,$ are arbitrary real constants, and
two constants  $\tilde \alpha $ and $\tilde \beta $ are given by
\be \tilde \alpha = -4\Delta (\sqrt{\alpha ^2+\beta^2}+2\alpha ),\
\tilde \beta =4\delta (\sqrt{\alpha ^2+\beta^2}-2\alpha ) , \ee
and four functions, $p_i$, $1\le i\le 4$, are defined by \bea
p_1(t)&=&\frac{2}{\Delta }f_{1,xx}|_{x=0}+2f_{1,x}|_{x=0}+\frac
{2\alpha}{\Delta } f_1|_{x=0}
-\frac{2\delta }{\Delta}f_{2,x}|_{x=0}-{4\delta } f_2|_{x=0}, \nonumber \\
p_2(t)& =&-\frac {2 \delta }{ \Delta } f_{1,x}|_{x=0}-{4\delta }
f_1|_{x=0}-
  \frac{2}{\Delta }f_{2,xx}|_{x=0}-{2 }f_{2,x}|_{x=0}-
\frac {2\alpha }{\Delta } f_2|_{x=0},\nonumber
\\
p_3(t)&=& -\frac{2}{\Delta
}f_{1,xx}|_{x=0}+2f_{1,x}|_{x=0}-\frac{2\alpha }{\Delta }
f_1|_{x=0} +\frac{2\delta }{\Delta}f_{2,x}|_{x=0}-{4\delta } f_2|_{x=0},\nonumber\\
p_4(t)&=&\frac{2\delta }{\Delta }f_{1,x}|_{x=0}-{4\delta }
f_1|_{x=0}+
  \frac{2}{\Delta }f_{2,xx}|_{x=0}-{2 }f_{2,x}|_{x=0}+
\frac{2\alpha }{\Delta } f_2|_{x=0}, \nonumber \eea with $\Delta $
and $\delta $ being defined by (\ref{eq:defofDeltaanddelta}).
\end{theorem}

\begin{proof} Note that now $f_1=f_1(x,t)$ and $f_2=f_2(x,t)$ are
functions of two variables $x$ and $t$. But in the solution
formula (\ref{eq:1compofsolofspecial2ndorderODEs:TAMS202}) and
(\ref{eq:3compofsolofspecial2ndorderODEs:TAMS202}), $t$ is the
dummy variable. Let us first compute that
\[\ba {rcl}
\phi_{1,t}^h&= &(D_{1,t}\cos \delta x +D_{2,t}\sin \delta
x)\textrm{e}^{\Delta x}+ (D_{3,t}\cos \delta x +D_{4,t}\sin \delta
x)\textrm{e}^{-\Delta x}\vspace{2mm}\\
\phi_{1,t}^s&=&-\D \int
_0^x\bigl[g_1(x-y)f_{1,t}(y,t)+g_2(x-y)f_{2,t}(y,t)\bigr]\,dy
\vspace{2mm}\\
&=&4\D \int
_0^x\bigl[g_1(x-y)f_{1,yyy}(y,t)+g_2(x-y)f_{2,yyy}(y,t)\bigr]\,dy
\vspace{2mm}\\
&=&
-4g_1(x)(f_{1,xx}|_{x=0})-4g_{1,x}(x)(f_{1,x}|_{x=0})-4g_{1,xx}(x)(f_1|_{x=0})
\vspace{2mm}
\\
&&-4g_2(x)(f_{2,xx}|_{x=0})-4g_{2,x}(x)(f_{2,x}|_{x=0})-4g_{2,xx}(x)(f_2|_{x=0})
-4\phi_{1,xxx}^s,
 \ea \]
where \[g_1=\frac 1 \Delta (\cos \delta x) (\sinh \Delta x),\
g_2=-\frac 1 \Delta (\sin \delta x)\,(\sinh \Delta x) .
\]
 Second, based on
(\ref{eq:phi1_x:TAMS202}) and (\ref{eq:phi1_xx:TAMS202}), we have
\[
\ba {l} \phi_{1,xxx}^h= \bigl[(\Delta \tilde {D}_{1}+\delta \tilde
{D}_{2})\cos \delta x +( \Delta \tilde {D}_{2}-\delta \tilde
{D}_{1})\sin \delta x\bigr]\textrm{e}^{\Delta x}
\vspace{2mm} \\
\qquad \qquad +\bigl[ (\delta \tilde {D}_4-\Delta \tilde
{D}_3)\cos \delta x -(\delta \tilde {D}_3+\Delta \tilde {D}_4)\sin
\delta x\bigr] \textrm{e}^{-\Delta x}, \ea \] and a direct
computation gives rise to
\[\ba {l}
g_{1,x}(x)=\D \frac {1}{2}(\cos \delta x)(\textrm{e}^{\Delta
x}+\textrm{e}^{-\Delta x})-\D \frac {\delta }{2\Delta }(\sin
\delta x)(\textrm{e}^{\Delta x}- \textrm{e}^{-\Delta x}),
\vspace{2mm}\\
g_{1,xx}(x)=(-\delta \sin \delta x+\D \frac{\alpha}{2\Delta }\cos
\delta x)\textrm{e}^{\Delta x} -(\delta \sin \delta x
+\D \frac {\alpha }{2\Delta }\cos \delta x)\textrm{e}^{-\Delta x}, \vspace{2mm}\\
g_{2,x}(x)=-\D \frac {1}{2}(\sin \delta x)(\textrm{e}^{\Delta
x}+\textrm{e}^{-\Delta x})-\D \frac {\delta }{2\Delta }(\cos
\delta x)(\textrm{e}^{\Delta x}- \textrm{e}^{-\Delta x}),
\vspace{2mm}\\
g_{2,xx}(x)=-(\delta \cos \delta x+\D \frac {\alpha }{2\Delta}
\sin \delta x)\textrm{e}^{\Delta x} -(\delta \cos \delta x -\D
\frac {\alpha }{ 2\Delta} \sin \delta x)\textrm{e}^{-\Delta x}.
 \ea
\]
Now a comparison of coefficients of four functionally independent
terms
$$(\cos \delta x)\textrm{e}^{\Delta x},\ (\sin \delta
x)\textrm{e}^{\Delta x},\  (\cos \delta x)\textrm{e}^{-\Delta x},\
(\sin \delta x)\textrm{e}^{-\Delta x}$$ in the equation
$\phi_{1,t}=-4\phi_{1,xxx}$, i.e., $\phi_{1,t}^h+\phi_{1,t}^s=
-4\phi_{1,xxx}^h-4\phi_{1,xxx}^s,$ yields that \bea
\label{eq:tpartlinearsystemofD1:TAMS202} D_{1,t}&=& -4(\Delta
\tilde {D}_1+\delta \tilde {D}_2)+p_1(t) \\ &=& 4(3\Delta \delta
^2-\Delta ^3)D_1-4(3\Delta ^2\delta -\delta ^3)D_2+p_1(t)
\nonumber \\ &=& \tilde {\alpha }D_1-\tilde {\beta
}D_2+p_1(t),\nonumber
\\ \label{eq:tpartlinearsystemofD2:TAMS202}
D_{2,t}&=& -4(\Delta \tilde {D}_2-\delta \tilde {D}_1)+p_2(t)
\\ &=& 4(3\Delta ^2\delta -\delta ^3)D_1+4 (3\Delta \delta
^2-\Delta ^3)D_2+p_2(t) \nonumber \\ &=& \tilde {\beta }D_1+\tilde
{\alpha }D_2+p_2(t),\nonumber
\\
D_{3,t}&=& -4(\delta \tilde {D}_4-\Delta \tilde {D}_3)+p_3(t)
 \\ &=& -4(3\Delta \delta^2 -\Delta ^3)D_3-4
(3\Delta ^2\delta -\delta ^3)D_4+p_3(t) \nonumber \\ &=&
-\tilde {\alpha }D_3-\tilde {\beta }D_4+p_3(t),  \nonumber \\
D_{4,t}&=& -4(-\delta \tilde {D}_3-\Delta \tilde {D}_4)+p_4(t)
\\ &=& 4(3\Delta ^2\delta -\delta ^3)D_3-4 (3\Delta \delta
^2-\Delta ^3)D_4+p_4(t) \nonumber \\ &=& \tilde {\beta }D_3-\tilde
{\alpha }D_4+p_4(t). \nonumber \eea These are two linear systems
of first-order ordinary differential equations. One is for $D_1$
and $D_2$ and the other is for $D_3$ and $D_4$. Solving these two
coupled linear systems for $D_i$, $1\le i\le 4$, we obtain the
general solution given by
(\ref{eq:1solofD_i1to4inthecaseoftildebetanot=0:TAMS202})  and
(\ref{eq:2solofD_i1to4inthecaseoftildebetanot=0:TAMS202}) in the
theorem. For example, we rewrite
(\ref{eq:tpartlinearsystemofD1:TAMS202}) and
(\ref{eq:tpartlinearsystemofD2:TAMS202}) as
\[
 \left[ \ba {c} D_{1,t}\vspace{2mm} \\ D_{2,t}\ea \right]=
\tilde {A}\left[ \ba {c} D_1\vspace{2mm} \\ D_2\ea \right]+\left[
\ba {c} p_1\vspace{2mm} \\ p_2\ea \right],\ \tilde {A}=\left[ \ba
{cc}\tilde {\alpha }&-\tilde {\beta }   \vspace{2mm} \\ \tilde
{\beta } &\tilde {\alpha }  \ea \right].
\]
Then it follows that
\[
\left[ \ba {c} D_1\vspace{2mm} \\ D_2\ea \right]=
\textrm{e}^{\tilde {A}t} \left[ \ba {c} D_{10}\vspace{2mm} \\
D_{20}\ea \right]+\textrm{e}^{\tilde {A}t}\int_0^t
\textrm{e}^{-\tilde {A}s}\left[ \ba {c} p_1(s)\vspace{2mm} \\
p_2(s)\ea \right]\,ds.
\]
Since $\tilde {A}$ has two complex eigenvalues
\[ \la ^\pm = \tilde {\alpha }\pm \tilde {\beta }\sqrt{-1}\,, \]
the evolution operator $\textrm{e}^{\tilde {A}t}$ reads as
\cite{HirshS-book1974,ArrowsmithP-book1992}
\[
\textrm{e}^{\tilde {A}t}= \textrm{e}^{ \tilde {\alpha }t}
\left[ \ba {cc} \cos \tilde{\beta  }t & -\sin \tilde{\beta  }t  \vspace{2mm} \\
\sin \tilde{\beta  }t & \cos \tilde{\beta  }t \ea \right].
\]
The derivation of the formulas for $D_3$ and $D_4$ is completely
similar. Note that the other equation $\phi_{2,t}=-4\phi_{2,xxx}$
is automatically satisfied, once $D_i,\ 1\le i\le 4$, are
determined as above. The proof is finished.
\end{proof}

\section{Wronskian solutions}
\setcounter{equation}{0}

In principle, we can construct general solutions associated with
two types of Jordan blocks of the coefficient matrix through the
solution formulas established in the previous section. In what
follows, we present a few special but interesting Wronskian
solutions, rather than general ones which are too complicated and
lengthy to present.

\subsection{Rational solutions}

Let us start with \begin{equation} \psi (\eta ) = \textrm{e}^{\eta
x-4\eta ^3t}- \textrm{e}^{-\eta x+4\eta ^3t}=2\,\textrm{sinh}\,
(\eta x -4\eta ^3t) ,
\end{equation}
where $\eta $ is a real constant. It is easy to see that this
function $\psi (\eta )$ solves
\begin{equation}
\psi _{xx}=\eta ^2\psi ,\ \psi _t=-4 \psi _{xxx}.
\label{eq:propertiessofpsi(eta):TAMS202}\end{equation} Upon
expanding \be \psi (\eta )= \sum _{i=0}^\infty \phi_i\eta
^{2i+1},\ee  from (\ref{eq:propertiessofpsi(eta):TAMS202}) we
obtain
\begin{equation}
\phi_{0,xx}=0,\ \phi_{i+1,xx}=\phi_{i},\
\phi_{i,t}=-4\phi_{i,xxx},\ i\ge 0.\end{equation}
 The
expressions of $\phi_i$ can be calculated through the series
expansion of the hyperbolic function $\sinh  x $:
\begin{equation}
\phi_i=\sum_{j=0}^{[(2i+1)/3]}\frac
{(-4)^j}{(2i-3j+1)!j!}x^{2i-3j+1}t^j,\ i\ge 0.
\end{equation}
Then for each $k\ge 1$, the associated Wronskian solution to the
KdV equation (\ref{eq:kdv:TAMS202}) is given by
\begin{equation}
u=-2\partial _x^2 \ln W(\phi_0,\phi_1,\cdots, \phi_{k-1}).
\end{equation}

Obviously, this solution is rational and it corresponds to the
following Jordan block:
\be  \left [\ba {cccc} 0 & & & 0\vspace{2mm}\\
1&0 & & \vspace{2mm}\\
& \ddots &\ddots & \vspace{2mm}\\
0 & & 1&0 \ea \right]_{k\times k}.\ee  We call it a rational
solution of order $k-1$. This type of solutions was also discussed
in
\cite{AblowitzS-JMP1978,SirianunpiboonHR-PLA1988,AdlerM-CMP1978}
and obtained by Freeman and Nimmo \cite{FreemanN-PLA1983} by
taking the limit $\eta _i \to 0$ of the multi-soliton solution.
Two rational solutions of lower-order are
\[\ba {l}u=-2\partial _x^2\ln W(\phi_0)=-2\partial _x^2 \ln W(x)=\D \frac 2
{x^2}, \vspace{2mm}\\ u=-2\partial _x^2 \ln
W(\phi_0,\phi_1)=-2\partial _x^2 \ln W(x,\D \frac 16 x^3-4t)=\D
\frac {6x^4-144xt}{(x^3+12t)^2}.\ea
\]

\subsection{Solitons, positons and negatons}

For a nonzero real eigenvalue $\lambda _i$, we start from the
eigenfunction $\phi_i(\lambda _i)$ determined by
\begin{equation}
 -(\phi_i (\lambda_i
))_{xx}=\lambda _i \phi_i(\lambda _i),\ (\phi_i(\lambda
_i))_{t}=-4(\phi_i(\lambda _i))_{xxx}.
\label{eq:oneofrepresentativeequations:TAMS202}\end{equation}
General solutions to this system in two cases of $\lambda _i>0$
and $\lambda _i<0$ read as
\begin{eqnarray}&&
 \phi_i(\lambda _i)=C_{1i}\sin (\eta _ix+4\eta _i^3t)+C_{2i}\cos (\eta
_ix+4\eta _i^3t),\ \eta _i=\sqrt{\lambda_i}\, ,
\label{eq:generalphiifornegativelambdai:TAMS202} \\
&&\phi_i(\lambda _i)=C_{1i} \sinh (\eta _ix-4\eta
_i^3t)+C_{2i}\cosh (\eta _ix-4\eta _i^3t),\ \eta
_i=\sqrt{-\lambda_i}\,
,\label{eq:generalphiiforpositivelambdai:TAMS202}
\end{eqnarray}
respectively, where $C_{1i}$ and $C_{2i}$ are arbitrary real
constants. By an inspection, we find that
 \begin{equation*}
-\left [\ba {c}\phi_i(\lambda _i) \vspace{2mm}\\
\frac 1{1!} \partial _{\lambda _i} \phi_i(\lambda _i )\vspace{2mm}\\
\vdots \vspace{2mm}\\
\frac 1{(k_i-1)!} \partial _{\lambda _i}^{k_i-1} \phi_i(\lambda _i
)
 \ea \right] _{xx}=
\left [\ba {cccc} \lambda_i  & & & 0\vspace{2mm}\\
1&\lambda_i  & & \vspace{2mm}\\
& \ddots &\ddots & \vspace{2mm}\\
0 & & 1&\lambda_i  \ea \right]_{k_i\times k_i}
 \left [\ba {c}\phi_i(\lambda _i) \vspace{2mm}\\
\frac 1{1!} \partial _{\lambda _i} \phi_i(\lambda _i )\vspace{2mm}\\
\vdots \vspace{2mm}\\
\frac 1{(k_i-1)!} \partial _{\lambda _i} ^{k_i-1}\phi_i(\lambda _i
)
 \ea \right] ,\end{equation*}
 and
 \begin{equation*}
( \frac 1{j!} \partial _{\lambda _i}^{j} \phi_i(\lambda _i )
 ) _{t}=-4 ( \frac 1{j!} \partial _{\lambda _i}^{j} \phi_i(\lambda _i )
) _{xxx},\ 0\le j\le k_i-1
 ,\end{equation*}
 where $\partial _{\lambda _i}$ denotes the derivative with respect
 to $\lambda _i$ and $k_i$ is an arbitrary non-negative integer.
Therefore, through this set of eigenfunctions, we obtain a
Wronskian solution to the KdV equation (\ref{eq:kdv:TAMS202}):
\begin{equation} u=-2\partial _x^2 \ln W(\phi_i (\lambda _i), \frac 1{1!} \partial _{\lambda
_i} \phi_i(\lambda _i) ,\cdots, \frac 1{(k_i-1)!}
\partial _{\lambda _i} ^{k_i-1}\phi_i (\lambda _i)),\end{equation}
which corresponds to the first type of Jordan blocks with a
nonzero real eigenvalue.

When $\lambda _i>0$, we get positon solutions
\cite{Matveev-PLA1992}, and when $\lambda _i<0$, we get negaton
solutions \cite{RasinariuSK-JPA1996}. A more general positon or
negaton
 can be obtained by combining $n$ sets of
eigenfunctions associated with different $\lambda _i>0$ or
different $\lambda _i<0$:
\begin{eqnarray} u&=&-2\partial _x^2 \ln W(\phi_1(\lambda _1) ,  \cdots, \frac 1{(k_1-1)!}
\partial _{\lambda _1} ^{k_1-1}\phi_1 (\lambda _1);\\
&& \cdots ; \phi_n (\lambda _n),  \cdots, \frac 1{(k_n-1)!}
\partial _{\lambda _n} ^{k_n-1}\phi_n (\lambda _n) ).
\nonumber
\end{eqnarray}
This solution is called an $n$-position or $n$-negaton of order
$(k_1-1,k_2-1,\cdots,k_n-1)$, to reflect the number of different
eigenvalues and the orders of derivatives of eigenfunctions. If
$k_i=1,$ $ 1\le i\le n$, we simply say that it is an $n$-position
or $n$-negaton.

An $n$-soliton solution is a special $n$-negaton: \begin{equation}
u=-2\partial _x^2 \ln (\phi_1,\phi_2,\cdots ,\phi _n)
\end{equation} with $\phi_i$ being given by
\begin{subequations}\begin{gather}
 \phi_i=\textrm{cosh}({\eta _ix-4\eta _i^3t}+\gamma_i) ,\  i \
 \textrm{odd},
\\
\phi_i=\textrm{sinh}({\eta _ix-4\eta _i^3t}+\gamma_i),\ \ i\
\textrm{even},
\end{gather}\end{subequations}
where $0<\eta _1<\eta _2<\cdots<\eta _n$ and $\gamma _i$, $1\le
i\le n$, are arbitrary real constants. There are other
representations of multi-soliton solutions to the KdV equation
(see, for example, \cite{Hirota-PRL1971,BryanS-NA1994}).
 Two kinds of special
positons of order $k$ are
\begin{eqnarray}&&
u=-2\partial _x^2\ln W(\phi,\partial _\lambda  \phi,\cdots,
\partial _\lambda ^{k-1}\phi ),\ \phi=\cos (\eta x+4\eta
^3t+\gamma(\eta ) ) ,\\ &&
 u=-2\partial _x^2\ln  W(\phi,\partial _\lambda  \phi,\cdots, \partial _\lambda ^{k-1}\phi
),\ \phi=\sin (\eta x+4\eta ^3t+\gamma(\eta ) ), \end{eqnarray}
 where $\lambda >0$, $\eta =\sqrt {\lambda }$ and $\gamma $ is an arbitrary function
 of $\eta$. But these two kinds of positons are equivalent to each
 other, due to the existence of the arbitrary function $\gamma$.
Similarly, two kinds of
 special negatons of order $k$ are
\begin{eqnarray}&&
u=-2\partial _x^2\ln  W(\phi,\partial _\lambda  \phi,\cdots,
\partial _\lambda ^{k-1}\phi ),\
  \phi=\textrm{cosh}({\eta x-4\eta ^3t+\gamma(\eta ) }) , \\ &&
 u=-2\partial _x^2\ln  W(\phi,\partial _\lambda  \phi,\cdots, \partial _\lambda ^{k-1}\phi
),\ \phi =\textrm{sinh}({\eta x-4\eta ^3t+\gamma(\eta ) }),
\end{eqnarray}
 where $\lambda <0$, $\eta =\sqrt {-\lambda }$ and $\gamma $ is an arbitrary function
 of $\eta$.
These solutions exhibited above were also discussed in a little
bit different way in
\cite{Arkad'evPP-ZNSLOMI1984,Matveev-PLA1992,RasinariuSK-JPA1996}.
The following are solitons, positons and negatons of lower-order:
\[\ba {rcl}u&=&-2\partial _x^2 \ln (\cosh (\eta x-4\eta^3 t+\gamma) )=
\D \frac {-2\eta^2}{\cosh ^2 (\eta x-4\eta^3 t+\gamma)},
\vspace{2mm}\\
u&=&-2\partial _x^2 \ln (\cos (\eta x+4\eta^3 t+\gamma) )= \D
\frac {2\eta^2}{\cos ^2 (\eta x+4\eta^3 t+\gamma)},\vspace{2mm}\\
u&=&-2\partial _x^2 \ln (\sinh (\eta x-4\eta^3 t+\gamma) )= \D
\frac {2\eta^2}{\sinh ^2 (\eta x-4\eta^3 t+\gamma)};
\vspace{2mm}\\
u&=&-2\partial _x^2 \ln W(\cosh (\eta_1 x-4\eta_1^3
t+\gamma_1), \sinh (\eta_2 x-4\eta_2^3 t+\gamma_2)  ) \vspace{2mm}\\
& =& \D \frac{4(\eta_1^2-\eta_2^2)[(\eta_2^2-\eta_1^2)+\eta
_1^2\cosh 2\theta _2 +\eta _2^2\cosh 2\theta _1 ]}{[(\eta _2-\eta
_1)\cosh (\theta _1+\theta _2) +(\eta _2+\eta _1)\cosh (\theta
_1-\theta _2) ]^2},
\vspace{2mm}\\
u&=&-2\partial _x^2 \ln W(\cos (\eta x+4\eta^3 t+\gamma),\partial
_\lambda \cos (\eta x+4\eta^3 t+\gamma)  ) \vspace{2mm}\\
& =& \D \frac {16 \eta^2 [2\cos ^2 (\eta x+4\eta^3 t+\gamma)+
(\eta x+12 \eta ^3t)\sin 2(\eta x+4\eta^3 t+\gamma) ] }{[2(\eta
x+12\eta ^3t)+ \sin 2 (\eta x+4\eta^3 t+\gamma)]^2},
\vspace{2mm}\\
u&=&-2\partial _x^2 \ln W(\cosh (\eta x-4\eta^3 t+\gamma),\partial
_\lambda \cosh (\eta x-4\eta^3 t+\gamma)  ) \vspace{2mm}\\
& =& \D \frac {16 \eta^2 [2\cosh ^2 (\eta x-4\eta^3 t+\gamma)-
(\eta x-12 \eta ^3t)\sinh 2(\eta x-4\eta^3 t+\gamma) ] }{[2(\eta
x-12\eta ^3t)+ \sinh 2 (\eta x-4\eta^3 t+\gamma)]^2};
 \ea \]
 where $\eta $, $\gamma $, $\eta _i$ and $\gamma_i$ are arbitrary real
 constants, and
 \[\theta _i=\eta _ix-4\eta _i^3t+\gamma _i,\ i=1,2.\]
Some parts of the graphs of these solutions with
$\gamma=\gamma_i=0$ are displayed in the following Figures 1 and
2.

\subsection{Complexitons}

For the second type of Jordan blocks of the coefficient matrix, we
start from a pair of eigenfunctions $\Phi_i(\alpha_i,\beta _i)
=(\phi_{i1}(\alpha_i,\beta _i),\phi_{i2}(\alpha_i,\beta _i))^T$
determined by
\begin{equation}
-\Phi_{i,xx}=A\Phi_i, \ \Phi_i=
\left[\ba {c} \phi_{i1}(\alpha_i,\beta _i) \vspace{2mm}\\
\phi_{i2}(\alpha_i,\beta _i) \ea \right] ,\ A_i=\left[\ba {cc}
\alpha
_i&-\beta _i\vspace{2mm}\\
\beta _i&\alpha _i\ea
\right],\label{eq:cond1ofJordantype2:TAMS202}\end{equation} and
\begin{equation}
 (\phi_{ij}(\alpha_i,\beta _i))_t =-4( \phi_{ij}(\alpha_i,\beta _i))_{xxx},\ j=1, 2.
 \label{eq:cond2ofJordantype2:TAMS202} \end{equation}
 The coefficient matrix $A_i$
 has two complex eigenfunctions: $\alpha _i\pm \beta
 _i\sqrt {-1}$.
By Theorem
\ref{thm:expressionofphi_1andphi_2ofcaseofcomplexeigenvalues:TAMS202},
a general solution to the homogenous system of
(\ref{eq:cond1ofJordantype2:TAMS202}) and
(\ref{eq:cond2ofJordantype2:TAMS202}) is \begin{eqnarray}
\phi_{i1}&=&\D \frac 12 [\cos (\delta_i (x-\bar \beta
_it)+\kappa_{1i})]\,\textrm{e}^{\Delta _i(x+\bar \alpha
_it)+\gamma _{1i} }  \\
&& + \D \frac 12 [\cos (\delta_i (x-\bar \beta
_it)+\kappa_{2i})]\,\textrm{e}^{-\Delta _i(x+\bar \alpha
_it)-\gamma _{2i} }\, ,\nonumber \\
 \phi_{i2}&=&\D \frac 12[ \sin
(\delta_i (x-\bar \beta _it)+\kappa_{1i})]\,\textrm{e}^{\Delta
_i(x+\bar \alpha _it)+\gamma _{1i} } \\
&& - \D \frac 12 [\sin (\delta_i (x-\bar \beta
_it)+\kappa_{2i})]\,\textrm{e}^{-\Delta _i(x+\bar \alpha
_it)-\gamma _{2i} }\, ,  \nonumber \end{eqnarray}
 where $\kappa_{ji}$ and
$\gamma_{ji}$ are arbitrary real constants and \bea && \Delta_i
=\sqrt{\frac {\sqrt{\alpha _i^2+\beta _i^2}-\alpha_i }{2}},\
\delta_i =\sqrt{\frac
{\sqrt{\alpha _i^2+\beta _i^2}+\alpha _i}{2}}, \label{eq:defofDeltaanddelta:TAMS202} \\
&&
 \bar {\alpha }_i=4\sqrt{\alpha_i ^2+\beta_i^2}+8\alpha _i
,\ \bar {\beta }_i=4\sqrt{\alpha_i ^2+\beta_i^2}-8\alpha_i .
\label{eq:defofbaralphaandbarbeta:TAMS202} \eea Similarly by an
inspection, we can see that
\begin{equation*}
-\left [\ba {c}\Phi_i \vspace{2mm}\\
\frac 1{1!} \partial _{\alpha  _i} \Phi_i  \vspace{2mm}\\
\vdots \vspace{2mm}\\
\frac 1{(l_i-1)!} \partial _{\alpha _i}^{l_i-1} \Phi_i
 \ea \right] _{xx}=
\left [\ba {cccc} A_i  & & & 0\vspace{2mm}\\
I_2&A_i  & & \vspace{2mm}\\
& \ddots &\ddots & \vspace{2mm}\\
0 & & I_2&A_i  \ea \right]_{l_i\times l_i} \left [
\ba {c}\Phi_i \vspace{2mm}\\
\frac 1{1!} \partial _{\alpha  _i} \Phi_i  \vspace{2mm}\\
\vdots \vspace{2mm}\\
\frac 1{(l_i-1)!} \partial _{\alpha _i}^{l_i-1} \Phi_i
 \ea \right]
,\end{equation*} and
\begin{equation*}
(\frac 1{j!} \partial _{\alpha  _i}^j \Phi_i  )_{t}=-4 (\frac
1{j!}
\partial _{\alpha  _i}^j \Phi_i  )_{xxx}, \ 0\le j\le l_i-1,
\end{equation*}
where $\partial _{\alpha _i}$ denotes the derivative with respect
to $\alpha_i$. Taking the derivative with respect to $\beta _i$,
we can have
\begin{equation*}
-\left [\ba {c}\Phi_i \vspace{2mm}\\
\frac 1{1!} \partial _{\beta  _i} \Phi_i  \vspace{2mm}\\
\vdots \vspace{2mm}\\
\frac 1{(l_i-1)!} \partial _{\beta _i}^{l_i-1} \Phi_i
 \ea \right] _{xx}=
\left [\ba {cccc} A_i  & & & 0\vspace{2mm}\\
\Sigma _2&A_i  & & \vspace{2mm}\\
& \ddots &\ddots & \vspace{2mm}\\
0 & & \Sigma _2&A_i  \ea \right]_{l_i\times l_i} \left [
\ba {c}\Phi_i \vspace{2mm}\\
\frac 1{1!} \partial _{\beta   _i} \Phi_i  \vspace{2mm}\\
\vdots \vspace{2mm}\\
\frac 1{(l_i-1)!} \partial _{\beta  _i}^{l_i-1} \Phi_i
 \ea \right]
\end{equation*}
 and
\begin{equation*}
(\frac 1{j!} \partial _{\beta  _i}^j \Phi_i  )_{t}=-4 (\frac 1{j!}
\partial _{\beta   _i}^j \Phi_i  )_{xxx}, \ 0\le j\le l_i-1,
\end{equation*}
where
\[ \Sigma_2=\left[\ba {cc} 0& -1 \vspace {2mm}\\
1&0 \ea \right] .\]
 Therefore, we obtain two Wronskian
 solutions to the KdV equation (\ref{eq:kdv:TAMS202}):
\begin{equation} u=-2\partial _x^2 \ln W(\Phi_i ^T, \frac 1{1!} \partial _{\alpha
_i} \Phi_i ^T,\cdots, \frac 1{(l_i-1)!}
\partial _{\alpha  _i} ^{l_i-1}\Phi_i ^T),\end{equation}
and \begin{equation} u=-2\partial _x^2 \ln W(\Phi_i ^T, \frac
1{1!} \partial _{\beta  _i} \Phi_i ^T,\cdots, \frac 1{(l_i-1)!}
\partial _{\beta  _i} ^{l_i-1}\Phi_i ^T),\end{equation}
which correspond to the same Jordan block of the second type but
are different. Such a general Wronskian solution is
\begin{equation} u=-2\partial _x^2 \ln W(\Phi_1 ^T, \cdots, \frac 1{(l_1-1)!}
\partial _{\zeta _1} ^{l_1-1}\Phi_1 ^T;\cdots ;
\Phi_n ^T, \cdots, \frac 1{(l_n-1)!}
\partial _{\zeta  _n} ^{l_n-1}\Phi_n ^T
),\end{equation} where $\partial _{\zeta _i}$ can be either of
$\partial _{\alpha _i}$ and $\partial _{\beta _i}$. This solution
is called an $n$-complexiton solution of order
$(l_1-1,l_2-1,\cdots,l_n-1)$, to reflect the orders of derivatives
of eigenfunctions with respect to eigenvalues. If $l_i=1,\ 1\le
i\le n$, we simply say that it is an $n$-complexiton.

In particular, one-complexiton  reads as \cite{Ma-PLA2002} \bea
\label{eq:onecomplexitonofKdV:TAMS202} u&=& -2\partial _x^2 \ln
W(\phi_{11},\phi_{12})
\\ &=& \D \frac {-4\beta _1^2\bigl[1+\cos(2\delta_1(x-\bar \beta
_1t)+2\kappa_{1})\cosh (2\Delta_1(x+\bar \alpha _1t)+2\gamma_{1})
\bigr]}{\bigl[\Delta _1 \sin (2\delta_1(x-\bar \beta
_1t)+2\kappa_{1})+\delta _1 \sinh (2\Delta_1(x+\bar \alpha
_1t)+2\gamma_{1})\bigr ]^2} \nonumber \\
\qquad && + \D \frac {4 \alpha _1\beta _1\sin (2\delta_1(x-\bar
\beta _1t)+2\kappa_{1})\sinh (2\Delta_1(x+\bar \alpha
_1t)+2\gamma_{1}) }{\bigl[\Delta _1 \sin (2\delta_1(x-\bar \beta
_1t)+2\kappa_{1})+\delta _1 \sinh (2\Delta_1(x+\bar \alpha
_1t)+2\gamma_{1})\bigr ]^2}, \nonumber
 \eea where $\alpha _1,\,\beta
_1>0,\, \kappa_1=\kappa_{11}=\kappa_{12},$ and
$\gamma_1=\gamma_{11}=\gamma_{12}$ are arbitrary real constants,
and $\Delta _1,\, \delta _1,\, \bar {\alpha }_1,$ and $\bar {\beta
}_1$ are given by
 \bea && \Delta_1 =\sqrt{\frac {\sqrt{\alpha _1^2+\beta
_1^2}-\alpha_1 }{2}},\ \delta_1 =\sqrt{\frac {\sqrt{\alpha
_1^2+\beta _1^2}+\alpha _1}{2}}, \nonumber  \\ &&  \bar {\alpha
}_1=4\sqrt{\alpha_1 ^2+\beta_1^2}+8\alpha _1 ,\ \bar {\beta
}_1=4\sqrt{\alpha_1^2+\beta_1^2}-8\alpha_1 . \nonumber \eea The
subcase of $\alpha _1=0$ leads to the following solution \[
u=\frac {8\beta _1+ 8\beta
_1\cos(\sqrt{2\beta_1}\,x-4\beta_1\sqrt{2\beta_1}\,t+2\kappa_{1})\cosh
(\sqrt{2\beta_1}\,x+ 4\beta_1\sqrt{2\beta_1}\, t+2\gamma_{1}) }
{\bigl[ \sin
(\sqrt{2\beta_1}\,x-4\beta_1\sqrt{2\beta_1}\,t+2\kappa_{1})+\sinh
(\sqrt{2\beta_1}\,x+4\beta_1\sqrt{2\beta_1}\,t+2\gamma_{1})\bigr
]^2}.\]
This solution is associated with purely imaginary eigenvalues of
the Schr\"odinger spectral problem with zero potential. In
addition, if we fix that \[ \kappa_1=\frac {\pi}4,\ \gamma_1=\frac
12 \ln (\frac ba ),\ \Delta _1=-a ,\ \delta _1=b,\] where $a$ and
$b$ are arbitrary real constants, then our one-complexiton
solution (\ref{eq:onecomplexitonofKdV:TAMS202}) boils down to the
breather-like or spike-like solution presented in
\cite{Jaworski-PLA1984}:
\[ u=\frac {8\{(a^2-b^2)(b/a)\cos \nu \sinh (\eta +p )+2b^2[1+\sin
\nu \cosh (\eta +p)]\}} {[\cos\nu -(b/a)\sinh (\eta +p)]^2},\]
where $\nu =-2bx+8(3a^2b-b^3)t,\, \eta =-2ax+8(a^3-3ab^2)t$ and $
p=\ln (b/a)$. Figure 3 
depicts some singularities of the single complexiton solution with
$\kappa=\gamma=0$.

\subsection{Interaction solutions}

We are now presenting examples of Wronskian interaction solutions
among different kinds of Wronskian solutions to the KdV equation.

Let us assume that there are two sets of eigenfunctions \be
\phi_1(\lambda ),\phi_2(\lambda ), \cdots ,\phi_k(\lambda );\
\psi_1(\mu ),\psi_2(\mu ), \cdots ,\psi_l(\mu
)\label{eq:twosetsofeigenfunctions:TAMS202}\ee associated with two
different eigenvalues $\lambda $ and $\mu $, respectively. A
Wronskian solution \be u=-2\partial _x^2 \ln W(\phi_1(\lambda
),\cdots,\phi_k(\lambda );\psi_1(\mu ),\cdots, \psi _l(\mu))\ee is
said to be a Wronskian interaction solution between two solutions
determined by the two sets of eigenfunctions in
(\ref{eq:twosetsofeigenfunctions:TAMS202}). Of course, we can have
more general Wronskian interaction solutions among three or more
kinds of solutions such as rational solutions, positons, solitons,
negatons, breathers and complexitons. Roughly speaking, it
increases the complexities of rational solutions, positons,
negatons and complexitons, respectively, to add zero, positive,
negative or complex eigenvalues to the spectrum of the coefficient
matrix.

In what follows, we would like to show a few special Wronskian
interaction solutions. Let us first choose different sets of
eigenfunctions:
\begin{eqnarray}
&& \phi_{\textrm{rational}}=cx+d, \ c,d=\textrm{consts.};\nonumber \\
&& \phi_{\textrm{soliton}}=\cosh (\eta x-4\eta ^3t+\gamma ),
\ \eta,\gamma =\textrm{consts.};\nonumber \\
&& \phi _{\textrm{positon}}=\cos (\eta x+4\eta ^3t+\gamma ),
\ \eta,\gamma =\textrm{consts.}; \nonumber \\
&& \phi _{\textrm{complexiton},1}=\cos (\frac
{\sqrt{2}}2(x-4t)+\kappa)\cosh ( \frac {\sqrt{2}}2(x+4t)+\gamma),\
\kappa,\gamma =\textrm{consts.},\nonumber
\\ &&
\phi _{\textrm{complexiton},2}= \sin (\frac
{\sqrt{2}}2(x-4t)+\kappa)\sinh ( \frac {\sqrt{2}}2(x+4t)+\gamma),\
\kappa,\gamma =\textrm{consts.}\nonumber
\end{eqnarray}
Three Wronskian interaction determinants between any two of a
rational solution, a single soliton and a single positon are
\begin{eqnarray}
&&W(\phi_{\textrm{rational}},\phi_{\textrm{soliton}})=\eta
(cx+d)\sinh (\eta x-4\eta ^3t +\gamma)-c\cosh (\eta x-4\eta ^3t
+\gamma ),\quad \quad \nonumber \\
&&W(\phi_{\textrm{rational}},\phi_{\textrm{positon}})=-\eta
(cx+d)\sin (\eta x +4\eta ^3 t+\gamma )-c\cos(\eta x +4\eta ^3
t+\gamma ),\nonumber
\\
&& W(\phi_{\textrm{soliton}},\phi_{\textrm{positon}})  = - \eta
\cosh (\eta x - 4\eta ^{3}t + \gamma )\sin (\eta x + 4\eta ^{3}t +
\gamma ) \nonumber \\&& \qquad \qquad \qquad \qquad \quad \ \  -
\eta \cos (\eta x + 4\eta ^{3}t + \gamma ) \sinh (\eta x - 4\eta
^{3}t + \gamma ). \nonumber  \eea Further the corresponding
Wronskian interaction solutions read as \[\ba{l}
 u_{rs} = -2\partial _x^2 \ln W(\phi_{\textrm{rational}},\phi_{\textrm{soliton}})
 = \D \frac {2\eta ^2[\eta ^2(cx+d)^2+c^2
\cosh ^2\xi_-  ]}{[\eta (cx+d) \sinh \xi_- - c \cosh \xi_- ]^2
},\vspace{2mm} \\
u _{rp}= -2\partial _x^2 \ln
W(\phi_{\textrm{rational}},\phi_{\textrm{positon}})
 =\D \frac {2\eta ^2[\eta ^2(cx+d)^2-c^2\cos ^2\xi _+ ]}{[\eta (cx+d)\sin \xi_+ +c\cos
\xi_+ ]^2},\vspace{2mm} \\
u_{sp}= -2\partial _x^2 \ln
W(\phi_{\textrm{soliton}},\phi_{\textrm{positon}})
 =\D \frac {4\eta ^{2}(\cosh ^2\xi_- + \cos ^2\xi_+ ) }
 {
 (\cosh \xi_- \sin \xi_+   + \cos \xi_+ \sinh \xi_-)
^{2}
 },\  \ea \]
where \[ \xi _{\pm} =\eta x \pm 4\eta ^{3}t + \gamma . \]
 Some singularities of these
solutions are shown in Figures 4 and 5.

The following are two Wronskian interaction determinants and
solutions involving three eigenfunctions. The first Wronskian
determinant is
\[\ba {rcl}
&&
W(\phi_{\textrm{rational}},\phi_{\textrm{soliton}},\phi_{\textrm{positon}})\vspace{2mm}\\
&=&  - \eta ^{2}[(cx + d)\eta (\cos \xi_+ \sinh \xi_- - \sin \xi_+
\cosh \xi_- ) - 2c\cos \xi_+ \cosh \xi_- ] \ea
\] so that its corresponding Wronskian solution reads as
\[ \ba {rcl}
u_{rsp}&= &-2\partial _x^2 \ln
W(\phi_{\textrm{rational}},\phi_{\textrm{soliton}},\phi_{\textrm{positon}})
\vspace{2mm}\\
& =&\D \frac{
  4(cx+d)\eta ^3(\sin \xi_+ \cosh \xi_-  +\cos
  \xi_+ \sinh \xi_-  )}
{(cx + d)\eta (\cos \xi_+ \sinh \xi_-  - \sin \xi_+ \cosh \xi_- )
- 2c\cos \xi_+ \cosh \xi_-  }
  \vspace{2mm} \\
&& +\D \frac {2\eta ^{ 2}[
 c \eta \cos \xi_+ \sinh \xi_-  + 2(cx + d)\eta \sin \xi_+ \sinh \xi_-  -
c\eta \sin \xi_+ \cosh \xi_- ]^{2}}
 { [ (cx + d)\eta (\cos \xi_+ \sinh \xi_-
 - \sin \xi_+ \cosh \xi_- ) - 2c\cos
\xi_+ \cosh \xi_- ]^2}, \ea \] where
\[ \xi_\pm =\eta x\pm 4\eta ^3 t +\gamma .\]
 The second Wronskian determinant and its
corresponding Wronskian solution are
\[\ba {rcl}
&&
W(x,\phi_{\textrm{complexiton,1}},\phi_{\textrm{complexiton,2}})\vspace{2mm}\\
&=&   - {\displaystyle \frac {\sqrt{2}}{4}} x \sin 2\xi
 + {\displaystyle \frac {\sqrt{2}}{4}} x\sinh 2\zeta
- \cosh ^2 \zeta  + \sin ^2 \xi   ,\ea
\] and
\[\ba {rcl}
u_{rc}&= &-2 \partial _x^2 \ln
W(x,\phi_{\textrm{complexiton,1}},\phi_{\textrm{complexiton,2}})
\vspace{2mm}\\
& =&
 [ \cosh ^4 \zeta -
\cos ^4 \xi  - 4 x^{2}\cos ^2\xi \cosh ^2 \zeta  +(2x^2-1)\cosh ^2 \zeta \vspace{2mm} \\
&&  +(2x^2+1) \cos ^2\xi  + 4\sqrt{2}\, x\cosh ^2\zeta \sin 2\xi +
4\sqrt{2}\, x\cos ^2\xi \sinh 2\zeta  \vspace{2mm} \\
&&  - \sqrt{2}\,x\sinh 2\zeta -\sqrt{2}\,x\sin 2\xi  -\D \frac 12
\sin 2\xi \sinh 2\zeta ]
 \left/ {\vrule height0.44em width0em depth0.44em} \right. \!
 \! (
 - {\displaystyle \frac {\sqrt{2}}{4}} x \sin 2\xi  \vspace{2mm} \\ &&
 + {\displaystyle \frac {\sqrt{2}}{4}} x\sinh 2\zeta
- \cosh ^2 \zeta + \sin ^2 \xi  )^2 , \ea \] where
 \[
\xi = {\displaystyle \frac {1}{2}} \,\sqrt{2}\,x - 2\, \sqrt{2}\,t
+ \kappa , \ \zeta = {\displaystyle \frac {1}{2}} \,\sqrt{2}\,x +
2\, \sqrt{2}\,t + \gamma .
\]
Figure 6 
depicts some singularities of these two solutions.

\section{Concluding remarks}

A broad set of sufficient conditions consisting of coupled systems
of linear partial differential equations has been presented, which
guarantees that the Wronskian determinant solves the Korteweg-de
Vries equation in the bilinear form. A systematical analysis has
been made for solving the resultant coupled systems of linear
partial differential equations, and solution formulas for their
representative systems have been explicitly presented. The key
technique is to apply the variation of parameters in solving the
involved non-homogeneous partial differential equations of
second-order and third-order. The whole analysis also offers an
approach to solve the coupled system of partial differential
equations:
\[ -\phi_{i,xx}=\sum_{i=1}^N\lambda _{ij}\phi_j ,\
\phi_{i,t}=-4\phi _{i,xxx},\ 1\le i\le N,\] where $\lambda _{ij}$
are arbitrary real constants.

Moreover, for each type of Jordan blocks of the coefficient matrix
$\Lambda =(\lambda _{ij})$, special sets of eigenfunctions have
been constructed and used to generate rational solutions,
solitons, positons, negatons, breathers, complexitons and their
interaction solutions to the Korteweg-de Vries equation. Of
course, the obtained solution formulas of the representative
systems allow us to construct more general Wronskian solutions
than rational solutions, positons, negatons, complexitons and
their interaction solutions presented in this paper. The resultant
solution structures also show us the rich diversity that the KdV
equation carries. We believe that any $(2+1)$-dimensional
counterpart of the KdV equation will have much more general
solution structures. This is because higher dimensional equations
have bigger spaces of initial data to choose.

However, any new explicit exact solutions, even Wronskian
solutions, to the KdV equation will still be very interesting. Two
open questions follow:
\begin{enumerate}
\item
  What about the following case? \[ \ba {l}
-\phi_{i,xx}=\D \sum_{j=1}^N\mu _{ij}\phi_{j,x}+\D
\sum_{j=1}^N\lambda _{ij}\phi_j,
\ 1\le i\le N,
\vspace{2mm}\\
\phi_{i,t}=\D \sum_{j=1}^N\zeta  _{ij}\phi_{j,xxx}+\D
\sum_{j=1}^N\xi  _{ij}\phi_j, \ 1\le i\le N,  \ea \] where all
coefficients are real constants. Are there any conditions, rather
than the ones established in this paper, which will guarantee the
Wronskian solutions for the KdV equation?

\item Are there any similar conditions to guarantee the double
Wronskian solutions for the KdV equation?
\end{enumerate}

On the other hand, it deserves more investigation whether there
exist Wronskian solutions for the generalized KdV equations (see,
say, \cite{Burroughs-Nonlinearity1993,You-CAM1996,SellY-book2002}
for examples) and what differential conditions on Wronskian
solutions one can have if there exist Wronskian solutions.

\section*{Acknowledgments} The authors would like to thank C. R. Gilson, K.
Manuro and M. Pavlov for stimulating discussions. They are also
grateful to the referee for helpful suggestions and valuable
comments. The work was supported in part by the University of
South Florida Internal Awards Program under Grant No. 1249-936RO.

\bibliographystyle{amsalpha}

\end{document}